\documentclass[12pt]{article}
\pdfoutput=1
\usepackage{hyperref}
\usepackage{graphicx}
\usepackage{graphics}
\usepackage{dsfont}
\usepackage{epsfig}
\usepackage{amsmath,amssymb,amsthm,amscd}
\usepackage{simpler-wick}
\usepackage {subcaption}

\newcommand{\Tr}{\textrm{Tr}}

\def\ket{\rangle}
\def\bra{\langle}

\newcommand{\be}{\begin{equation}}
\newcommand{\ee}{\end{equation}}
\newcommand{\ba}{\begin{aligned}}
\newcommand{\ea}{\end{aligned}}

\numberwithin{equation}{section}

\begin{document}
\begin{titlepage}

\rightline{USTC-ICTS/PCFT-22-32}

\vskip 3 cm

\centerline{\Large 
\bf  
Krylov Complexity    } 
\vskip 0.2 cm
\centerline{\Large 
\bf  
 in Calabi-Yau Quantum Mechanics     }

\vskip 0.5 cm

\renewcommand{\thefootnote}{\fnsymbol{footnote}}
\vskip 30pt \centerline{ {\large \rm 
Bao-ning Du\footnote{baoningd@mail.ustc.edu.cn},  Min-xin Huang\footnote{minxin@ustc.edu.cn}
} } \vskip .5cm  \vskip 20pt 

\begin{center}
{Interdisciplinary Center for Theoretical Study,  \\ \vskip 0.1cm  University of Science and Technology of China,  Hefei, Anhui 230026, China} 
 \\ \vskip 0.3 cm
{Peng Huanwu Center for Fundamental Theory,  \\ \vskip 0.1cm  Hefei, Anhui 230026, China} 
\end{center}

\setcounter{footnote}{0}
\renewcommand{\thefootnote}{\arabic{footnote}}
\vskip 40pt
\begin{abstract}

Recently, a novel measure for the complexity of operator growth is proposed based on  Lanczos algorithm and Krylov recursion method. We study this Krylov complexity in quantum mechanical systems derived from some well-known local toric Calabi-Yau geometries, as well as some non-relativistic models. We find that for the Calabi-Yau models, the Lanczos coefficients grow slower than linearly for small $n$'s, consistent with the behavior of integrable models. On the other hand, for the non-relativistic models, the Lanczos coefficients initially grow linearly for small $n$'s, then reach a plateau. Although this looks like the behavior of a chaotic system, it is mostly likely due to saddle-dominated scrambling effects instead, as argued in the literature. In our cases, the slopes of linearly growing Lanczos coefficients almost saturate a bound by the temperature.  During our study, we also provide an alternative general derivation of the bound for the slope. 

\end{abstract}

\end{titlepage}
\vfill \eject


\newpage

\baselineskip=16pt

\tableofcontents

\section{Introduction}
\label{sectionintro}

The understandings of complexity and chaos are important in both condensed matter physics and high energy physics, see e.g. \cite{Maldacena:2015waa, Susskind:2018pmk, BalasubramanianTalk}.  Recently, a novel measure for the complexity of operator growth is proposed based on Lanczos algorithm and Krylov recursion method \cite{Parker:2018yvk}, where it is hypothesized that the asymptotic growth behavior can distinguish between integrable or chaotic quantum systems. Known as the Krylov complexity, it has been studied extensively in the literature, e.g.  \cite{Jian:2020qpp, Rabinovici:2020ryf,  Rabinovici:2022beu, Caputa:2021sib, Balasubramanian:2022tpr,  Balasubramanian:2022dnj}.  For studies of other measures of complexity as a useful indicator of integrable or chaotic nature of the underlying quantum system, see e.g. the recent papers \cite{Balasubramanian:2019wgd, Balasubramanian:2021mxo}.

In this paper, we compute the Lanczos coefficients and Krylov complexity for some one-dimensional one-body quantum systems derived from some well-known local toric Calabi-Yau geometries, studied recently in the context a bootstrap method in \cite{Du:2021hfw}. The Hamiltonians of these models involve exponential functions of the position and momentum operators, and may provide grounds for exploring novel phenomena not available in conventional quantum systems. These models are not only purely of mathematical interest. Indeed, they can describe some experimentally confirmed phenomena in condensed matter physics such as the Hofstadter's butterfly, see e.g. \cite{Hatsuda:2016mdw, Hatsuda:2020ocr}.  For comparison purpose, we also study some non-relativistic models with the standard quadratic kinetic term for the momentum operator, including the the well-known quartic anharmonic oscillator model and a Toda model. 

These models are expected to be integrable. For examples, for the Calabi-Yau models, the exact quantitated conditions including non-perturbative contributions were proposed in terms of topological string theory in \cite{Grassi:2014zfa, Wang:2015wdy}.  The case of local $\mathbb{P}^1\times \mathbb{P}^1$ Calabi-Yau model also belongs to the class of integrable systems known as the relativistic Toda models. We shall test the asymptotic growth hypothesis proposed in \cite{Parker:2018yvk}. We find that the Lanczos coefficients of the Calabi-Yau models do grow slower than linearly, consistent with the behavior of integrable models. In our numerical calculations, we choose two different initial operators $\mathcal{O}_0=\hat{x},e^{\hat{x}}$, and find that their Lanczos coefficients have almost the same shape with somewhat smoother behavior for the case of $\mathcal{O}_0=e^{\hat{x}}$.

For the quantum chaotic systems, the Lanczos coefficients grow linearly $b_n\sim\alpha n$, and it is argued in \cite{Parker:2018yvk} that the slope $\alpha$ satisfy a bound by the temperature
\be  \label{bound3.3}
\alpha \leq \pi T,
\ee
which is related to the famous bound on Lyapunov exponents of chaos in \cite{Maldacena:2015waa}. We use the convergence property of the autocorrelation function to provide an alternative probably simpler derivation of the bound.

For the non-relativistic models, we find that the initial growth of $b_n$ is linear. The similar phenomenon also occurs in some other contexts due to the saddle-dominated scrambling \cite{Xu:2019lhc,Bhattacharyya:2020art}, and is not necessarily a decisive indicator of chaotic behavior. A feature of the non-relativistic models is that the slopes is proportional to the temperature and  the bound (\ref{bound3.3}) is  always almost saturated, similar as the inverted harmonic oscillator in \cite{Baek:2022pkt}. 

Some other related recent studies include the Krylov complexity in dissipative open quantum systems \cite{Bhattacharya:2022gbz, Bhattacharjee:2022lzy}, the applications to topics such as Bondi-Metzner-Sachs symmetry and flat space holography \cite{Banerjee:2022ime},  various aspects of cosmology \cite{Haque:2021hyw, Adhikari:2022oxr}  and quantum field theory \cite{Adhikari:2022whf}. 

The paper is organized as the followings. In section \ref{review}, we review the formalism of Krylov complexity and  use the harmonic oscillator as a warm-up example. In section \ref{derivebound}, we provide a derivation of the  bound for the slope of linearly growing Lanczos coefficients.  In section \ref{CY}, we study the Calabi-Yau models. In sections \ref{quartic} and \ref{Toda}, we consider the quartic anharmonic oscillator and a non-relativistic Toda model respectively. Finally in section \ref{discussions} we finish with some discussions.

\section{Review of the formalism} 
\label{review}

Consider a quantum system with the Hamiltonian $H$. An operator $\mathcal{O}$ can be regarded as a state 
in the Hilbert space of operators, denoted as $|\mathcal{O} )$ to distinguish from the bra-ket notation in the usual Hilbert space.  A Wightman inner product, considered e.g. in \cite{Jian:2020qpp, Guo:2022hui}, can be defined on the Hilbert space of operators
\be \label{Wightman}
 (\mathcal{O}_1 | \mathcal{O}_2 ) = \frac{\Tr (e^{-\frac{\beta H}{2}} \mathcal{O}_1^\dagger e^{-\frac{\beta H}{2}}  \mathcal{O}_2 ) }
 {\Tr e^{- \beta H}}, 
\ee
where $\beta=\frac{1}{T}$ is the inverse temperature in our convention. The definition is properly normalized by the partition function $Z=\Tr e^{- \beta H}$. We can define the norm of an operator $||\mathcal{O} || := ( \mathcal{O} | \mathcal{O} )^{\frac{1}{2}} $ due to the positivity of inner product. 

In the Heisenberg picture of quantum mechanics, the evolution of an operator can be described by the Liouvillian action 
\be 
\mathcal{L} \mathcal{O} := [H, \mathcal{O}]. 
\ee 
Starting with an operator $|\mathcal{O})$, we can apply the Gram-Schmidt procedure to orthogonalize the sequence $\{ \mathcal{L}^n |\mathcal{O}) \}$.  This is known as the Lanczos algorithm. It is conventional to start with a Hermitian operator  $|\mathcal{O}_0)$ with proper normalization $||\mathcal{O}_0||=1$. Then we have the next normalized operator  $|\mathcal{O}_1):=b_1^{-1} \mathcal{L}|\mathcal{O}_0) $ where $b_1 := || \mathcal{L} \mathcal{O} ||$. For $n\geq 2$, the recursion relation is 
\be \ba \label{recursion}
& |A_n) := \mathcal{L} |\mathcal{O}_{n-1} ) - b_{n-1} | \mathcal{O}_{n-2}) , \\ 
& b_n := || A_n||, ~~~~~  |\mathcal{O}_n) := b_n^{-1} |A_n). 
\ea \ee
The recursion may terminate at a finite step if $b_n=0$. It is straightforward to show inductively that the operator $i^n \mathcal{O}_n$ is Hermitian, so $(\mathcal{O}_n| \mathcal{L} |\mathcal{O}_n) =0$. We have an orthonormal sequence $|\mathcal{O}_n) , n=0,1,2, \cdots$ under the Wightman inner product, known as the Krylov basis. The autocorrelation function is defined as 
\be
C(t) = ( \mathcal{O}_0 | e^{i \mathcal{L} t} | \mathcal{O}_0 ) .
\ee

The sequence of numbers $b_n$ is known as the Lanczos coefficients. It is expected that they initially grow in a ramping region and then eventually reach a plateau. It is shown that the growth is at most linear \cite{Parker:2018yvk}, and hypothesized that this is saturated $b_n\sim\alpha n$ by chaotic quantum models, while the growth behavior of integrable models is slower 
\be
b_n\sim n^{\delta}, ~~~~ 0<\delta <1. 
\ee

Now we consider the time evolution of an operator $\mathcal{O}(t) =e^{iHt} \mathcal{O}_0  e^{-iHt}$ in the Heisenberg picture, and expand it in the orthonormal Krylov basis as
	\be
		|\mathcal{O}(t))=\sum_{n=0}^{\infty} i^n \phi_n(t) |\mathcal{O}_n),
	\ee
where the coefficient $\phi_n(t)$ can be expressed as
	\be \label{phi2.7}
		\phi_n(t)=i^{-n}(\mathcal{O}_n|\mathcal{O}(t)). 
	\ee
This means that $\phi_n(t)$ can be regarded as the wave function of the operator, satisfying the normalization condition
	\be
		\sum_{n=0}^{\infty} |\phi_n(t)|^2=1.
	\ee 

Using the Heisenberg time evolution equation, we can get the differential equations
	\be
		\partial_t\phi_n = -b_{n+1}\phi_{n+1} + b_n\phi_{n-1},~~~\phi_n(0)=\delta_{n0},~\phi_{-1}(t)=0.
	\ee  
This equation tell us that all $\phi_n(t)(n>1)$ can be solved recursively with the initial condition $\phi_0(t)=C(t)$ which is easily derived from the Baker-Campbell-Hausdorff formula. Finally, the Krylov complexity is defined as 
	\be \label{Krylov2.10}
		K(t):=\sum_n n|\phi_n(t)|^2.
	\ee
For the integrable systems, in the early time, it is expected that the Krylov complexity satisfies polynomial growth:
	\be
		K(t)\sim t^{\frac{1}{1-\delta}},
	\ee 
while for the chaotic systems, it has an exponential growth  
	\be
		K(t)\sim e^{2\alpha t}. 
	\ee
The parameters $\delta$ and $\alpha$ are the same as appearing the the growth of the Lanczos coefficients. 	

As a warm-up exercise, it is illustrative to work out the  Lanczos algorithm for the simple case of harmonic oscillator
\be 
H = \frac{1}{2} ( \hat{x}^2  +\hat{p}^2) , 
\ee 
with the canonical commutation relation $[\hat{x}, \hat{p} ]=i\hbar$.   We can compute the norm of the operator $\hat{x}$
\be
\Tr (e^{-\frac{\beta H}{2} } \hat{x}  e^{-\frac{\beta H}{2} } \hat{x} ) = \sum_{m,n=0}^{\infty} e^{-\frac{\beta}{2} (E_m+E_n)} |\bra m | \hat{x} |n\ket |^2 ,
\ee
where we insert the energy eigenstates and $E_n =(n+\frac{1}{2}) \hbar$. The position operator can be written in terms of the creation and annihilation operators $\hat{x} =\sqrt{\frac{\hbar}{2}} (\hat{a} + \hat{a}^\dagger)$, so the sum is only non-vanishing for the cases $m=n\pm 1$ and can be easily done. The calculations for the momentum operator $\hat{p}$ are similar. Including the normalization from the partition function  $\Tr(e^{-\beta H}) =\frac{1}{2\sinh (\frac{\beta \hbar}{2}) } $, we find 
\be 
||\hat{x} ||^2 = ||\hat{p} ||^2 = \frac{\hbar}{ 2 \sinh(\frac{\beta\hbar}{2}) } .  
\ee 

We can apply the Lanczos algorithm starting from a normalized operator 
\be
\mathcal{O}_0 =  \sqrt{ \frac{ 2 \sinh(\frac{\beta\hbar}{2}) }{\hbar} } \hat{x} . 
\ee
The first recursion gives 
\be
\mathcal{O}_1 =  -i \sqrt{ \frac{ 2 \sinh(\frac{\beta\hbar}{2}) }{\hbar} } \hat{p} , ~~~ b_1=\hbar . 
\ee
Then the second operator in the recursion (\ref{recursion}) actually vanishes $A_2 =0, b_2=0$. So the harmonic oscillator is a rather trivial model with the Lanczos algorithm terminating at the second step. 

There is a simple algorithm to compute the Lanczos coefficients from the {\it moments}, explained e.g. in \cite{Parker:2018yvk}. The {\it moments} are defined in terms of the initial properly normalized operator as 
\be 
\mu_{n} := (\mathcal{O}_0 | \mathcal{L}^{n} | \mathcal{O}_0 ) , 
\ee
which is vanishing for odd $n$, since the operator is Hermitian.  Then (correcting a typo in \cite{Parker:2018yvk}) we have the relation 
\be \label{relation2.11}
b_1^{2n}b_2^{2(n-1)} \cdots b_n^2 = \det(\mu_{i+j}) _{0\leq i,j \leq n} . 
\ee
So the Lanczos coefficients can be recursively efficiently computed. For example, in the above harmonic oscillator example, it is easy to compute $\mu_{2n}=\hbar^{2n}$ for any $n\geq 0$. Therefore the  Lanczos coefficients  are $b_1=\hbar, b_2=0$.   

As a slight generalization, consider a quantum system of $N$ decoupled harmonic oscillators of different frequencies 
\be 
H = \sum_{i=1}^N \frac{1}{2} (\hat{p}_i^2 +\omega_i^2 \hat{x}_i^2) ,
\ee 
with a properly normalized initial operator $\mathcal{O}_0 \sim \sum_{i=1}^N \hat{x}_i$.  It is easy to compute the moments in this case 
$\mu_{n} = \frac{\hbar^{n}  \sum_{i=1}^N \omega_i^n}{N}$ for even $n$. The determinant in (\ref{relation2.11}) vanishes for $n\geq 2N$, so the Lanczos algorithm terminates at the $2N$ step with $b_{2N}=0$.

\section{A derivation of the bound for the slope of linearly growing Lanczos coefficients} 
\label{derivebound}

Before studying concrete models, in this section let us provide another general discussion relating to the various bounds  on quantum chaos.  In the seminal paper \cite{Maldacena:2015waa}, a universal bound for the Lyapunov exponent $\lambda_L$
 of  quantum chaos is proposed based on holographic arguments
 \be  \label{bound3.1}
 \lambda_L \leq 2\pi T .
 \ee
The Lyapunov exponent can be computed from the four-point OTOC's (out-of-time-order correlators), which have been subsequently studied extensively in the literature, e.g. in a recent paper \cite{Hashimoto:2020xfr}. On the other hand, for quantum chaotic systems with  linearly growing Lanczos coefficients $b_n\sim \alpha n$, it is proved for infinite temperature and conjectured with strong evidence for finite temperature in \cite{Parker:2018yvk} that the Lyapunov exponent is bound by the slope 
\be   \label{bound3.2}
 \lambda_L \leq 2\alpha . 
 \ee
By analyzing the asymptotic behavior of the spectral function, it is also argued in \cite{Parker:2018yvk}  that the slope is bound by (\ref{bound3.3}). Therefore the bound (\ref{bound3.2}) from Lanczos coefficients is generally stronger than the earlier result (\ref{bound3.1}). 

Here we provide an alternative derivation of the bound (\ref{bound3.3}) without using the spectral function.  Let's define a new function
	\be \ba
		b(\gamma,\epsilon):&=\frac{1}{2Z}\{ \Tr ( e^{(-\frac{\beta}{2}+\gamma)\hat{H}}\mathcal{O}^\dagger e^{(-\frac{\beta}{2}+\epsilon)\hat{H}}\mathcal{O}  ) +\Tr ( e^{(-\frac{\beta}{2}-\gamma)\hat{H}}\mathcal{O}^\dagger e^{(-\frac{\beta}{2}-\epsilon)\hat{H}}\mathcal{O}  )   \}\\
		&= \frac{1}{Z} \sum_{k=0}^\infty\sum_{m+n=2k} \frac{\gamma^m\epsilon^n}{m!n!}\Tr ( e^{-\frac{\beta}{2}\hat{H}} \hat{H}^m \mathcal{O}^\dagger e^{-\frac{\beta}{2}\hat{H}} \hat{H}^n \mathcal{O} ),
	\ea\ee
where $\mathcal{O} = \mathcal{O}_0$ is the initial operator in a Lanczos algorithm. 
	
The moments can also be expressed as
	\be\ba
		\mu_{2n}= \frac{1}{Z} \sum_{i+j=2n} \frac{(2n)!}{i!j!}(-1)^i \Tr (  e^{-\frac{\beta}{2}\hat{H}} \hat{H}^i \mathcal{O}^\dagger e^{-\frac{\beta}{2}\hat{H}} \hat{H}^j \mathcal{O}  ).
	\ea\ee
If we set the variable $\epsilon =- \gamma$, we have the following relation
	\be \ba \label{bgamma}
		b(\gamma,-\gamma)&=\frac{1}{Z}\Tr ( e^{(-\frac{\beta}{2}+\gamma)\hat{H}}\mathcal{O}^\dagger e^{(-\frac{\beta}{2}-\gamma)\hat{H}}\mathcal{O}  )\\
		& =\sum_{n=0}^{\infty} \frac{\gamma^{2n}}{(2n)!} \mu_{2n}\\
		& = ( \mathcal{O} | e^{\mathcal{L}\gamma} | \mathcal{O} ).
	\ea\ee
For $\gamma=0$, $b(0,0)=( \mathcal{O} | \mathcal{O} )_{\beta}$ is the norm square of the operator $\mathcal{O}$, where we denote the temperature parameter $\beta$ explicitly. Since $b(\gamma,-\gamma)=b(-\gamma,\gamma)$, we can for convenience assume that $\gamma \geq 0$. And on the third equality of the above equation, we use the fact that $\mu_{2n+1}=0$. So we find that this is actually the autocorrelation function $b(\gamma,-\gamma)=C( -i\gamma)$.

We consider the radius of convergence of $b(\gamma,-\gamma)$ as a function of $\gamma$. We assume that the energy eigenvalues are non-negative and the norm square $( \mathcal{O} | \mathcal{O} )_{\beta}$ is finite for any finite temperature $\beta>0$ with the choice of the initial operator $\mathcal{O}$. Then we can estimate 
	\be\ba
		b(\gamma,-\gamma)&=\frac{1}{Z}\Tr ( e^{(-\frac{\beta}{2}+\gamma)\hat{H}}\mathcal{O}^\dagger e^{(-\frac{\beta}{2}-\gamma)\hat{H}}\mathcal{O}  )\\
		& = \frac{1}{Z}\sum_{m,n} e^{(-\frac{\beta}{2}+\gamma)E_m+(-\frac{\beta}{2}-\gamma)E_n}|\langle m|\hat{\mathcal{O}}|n\rangle|^2\\
		& < \frac{1}{Z}\sum_{m,n} e^{(-\frac{\beta}{2}+\gamma)E_m+(-\frac{\beta}{2}+\gamma)E_n}|\langle m|\hat{\mathcal{O}}|n\rangle|^2\\
		&=\frac{1}{Z}\Tr ( e^{(-\frac{\beta}{2}+\gamma)\hat{H}}\mathcal{O}^\dagger e^{(-\frac{\beta}{2}+\gamma)\hat{H}}\mathcal{O}  )\\
		&=(\mathcal{O}|\mathcal{O})_{\beta-2\gamma}.
	\ea\ee
So this is finite  when $\gamma<\frac{\beta}{2}$, and the radius of convergence is no less than $\frac{\beta}{2}$. Of course the true radius of convergence can be much bigger than $\frac{\beta}{2}$. In fact for the case of simple harmonic oscillator as reviewed in section \ref{review}, it is easy to check that $b(\gamma,-\gamma)$ is finite for any $\gamma$, so the radius of convergence is infinite in this simple case.

On the other hand, as explained in \cite{Parker:2018yvk}, for the simple example of $b_n=\alpha n$, the autocorrelation function can be computed exactly $C(t) = \frac{1}{\cosh (\alpha t)} $, while the moments are known as Euler or secant numbers  and have the asymptotic behavior 
\be
\mu_{2n} \sim (\frac{4 n\alpha}{e\pi})^{2n}. 
\ee 
For a general linearly growing Lanczos coefficients $b_n \sim \alpha n$ with some possible small deviations from the exact case, it is expected and checked in some examples that the above asymptotic behavior is robust and still valid \cite{Parker:2018yvk}. Since $\cosh (\alpha t)$ has a zero at $t=\frac{\pi i}{2\alpha}$, the radius of convergence is  $\frac{\pi }{2\alpha}$. Combining the arguments in the previous paragraph we arrive at the bound (\ref{bound3.3}).

\section{Calabi-Yau models} 
\label{CY}

We consider the local $\mathbb{P}^1\times\mathbb{P}^1$ model and the local $\mathbb{P}^2$ model. Their Hamiltonians are 
	\be\ba  \label{CYHamiltonian}
		e^{\hat{H}}&=e^{\hat{x}}+e^{-\hat{x}}+e^{\hat{p}}+e^{-\hat{p}},\\
		e^{\hat{H}}&=e^{\hat{x}}+e^{\hat{p}}+e^{-\hat{x}-\hat{p}},\\
	\ea\ee
where the position and momentum operators satisfy the canonical commutation relation $[\hat{x}, \hat{p}]=i\hbar$. The energy eigenvalues and eigenstates can be computed numerically by truncating to a finite level in a basis of eigenfunctions of a simple harmonic oscillator, as implemented in \cite{Huang:2014eha}. 

The partition function can be expressed as $Z=\Tr e^{-\beta \hat{H}}=\Tr \hat{\rho}^{\beta} $, where $\hat{\rho}:= e^{- \hat{H}}$ is of trace class and positive definite.   In some Calabi-Yau models, there are simple formulas for the operator $\hat{\rho}$ in the basis of position eigenstates. For the example, the formula for the $\mathbb{P}^1\times \mathbb{P}^1$ model is known in the literature, can be found e.g. in our previous paper \cite{Du:2020nwl} 
\be
		\hat{\rho}(x_1,x_2)=\langle x_1|\hat{\rho}|x_2\rangle =\frac{u(x_1)^{-\frac{1}{2}} u(x_2)^{-\frac{1}{2}}}{4\hbar \cosh (\frac{\pi(x_1-x_2)}{2\hbar})},
	\ee 
where $u(x)= 2\cosh( \frac{x}{2} )$. For the local $\mathbb{P}^2$ model, the formula is more complicated and involves Faddeev's quantum dilogarithm. An analytic integral formula the partition function can be obtained when $\beta$ is an integer.  In this case, we have 
\be \label{partitionfunction}
		Z=\Tr e^{-\beta H}=\int dx_1 \int dx_2\dots \int dx_\beta \hat{\rho}(x_1,x_2)\hat{\rho}(x_2,x_3)\dots \hat{\rho}(x_\beta,x_1).
	\ee
However the Lanczos algorithm is difficult to implement analytically even for integer $\beta$, since the Hamiltonians involve logarithm. One possible approach is to replace the definition of Hamiltonians in (\ref{CYHamiltonian}) with $e^{\hat{H}}\rightarrow \hat{H}$ and correspondingly use a modified  Wightman inner product  by replacing the  $e^{-\beta \hat{H}}\rightarrow \hat{H}^{-\beta}$ in (\ref{Wightman}). This appears to be a consistent definition since the  modified Wightman inner product would be still positive definite.

\begin{figure}
	\centering
	\begin{subfigure}{0.49\textwidth}
		\includegraphics[width=\textwidth]{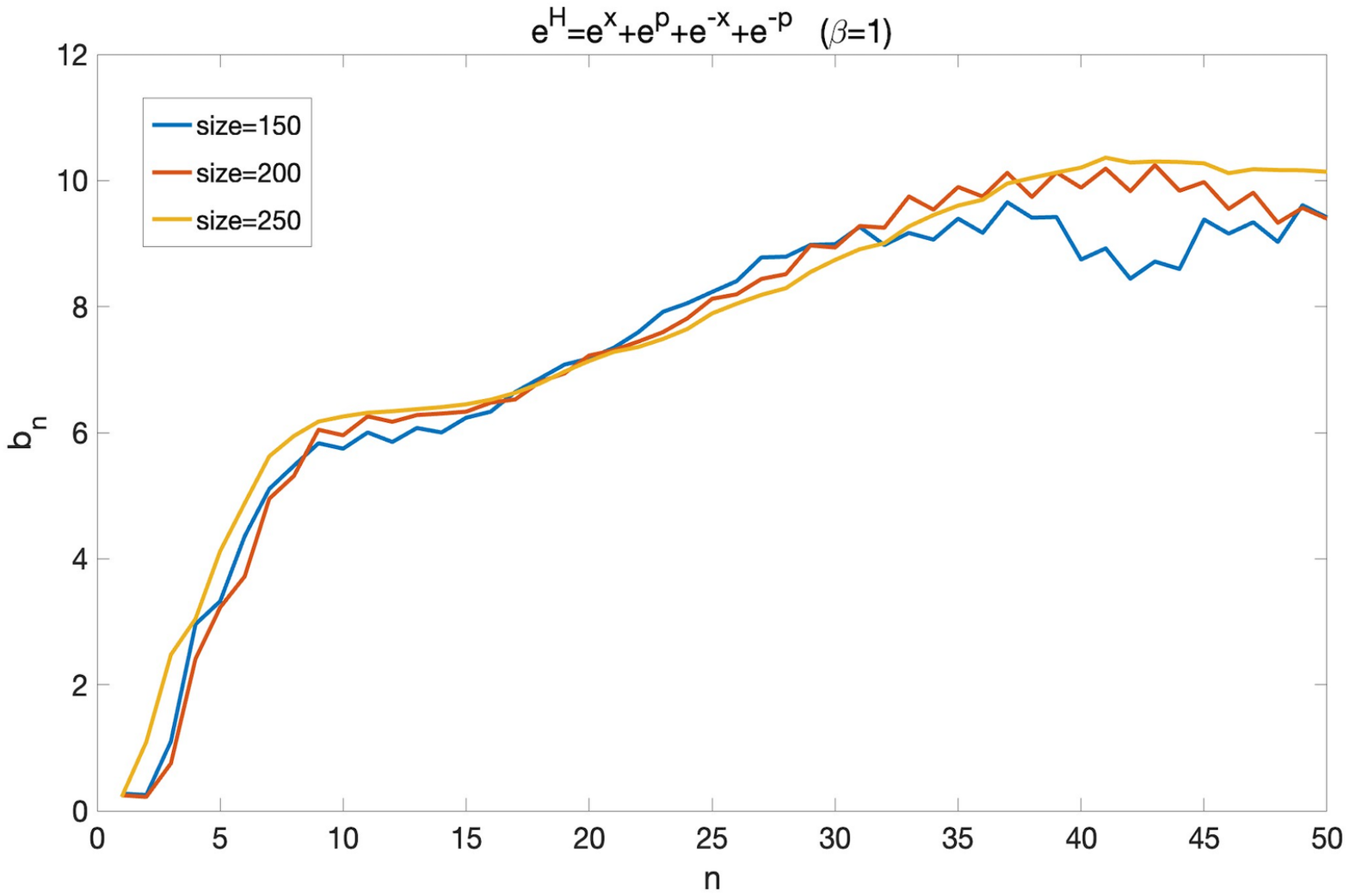} 	
	\end{subfigure}
	\begin{subfigure}{0.49\textwidth}
		\includegraphics[width=\textwidth]{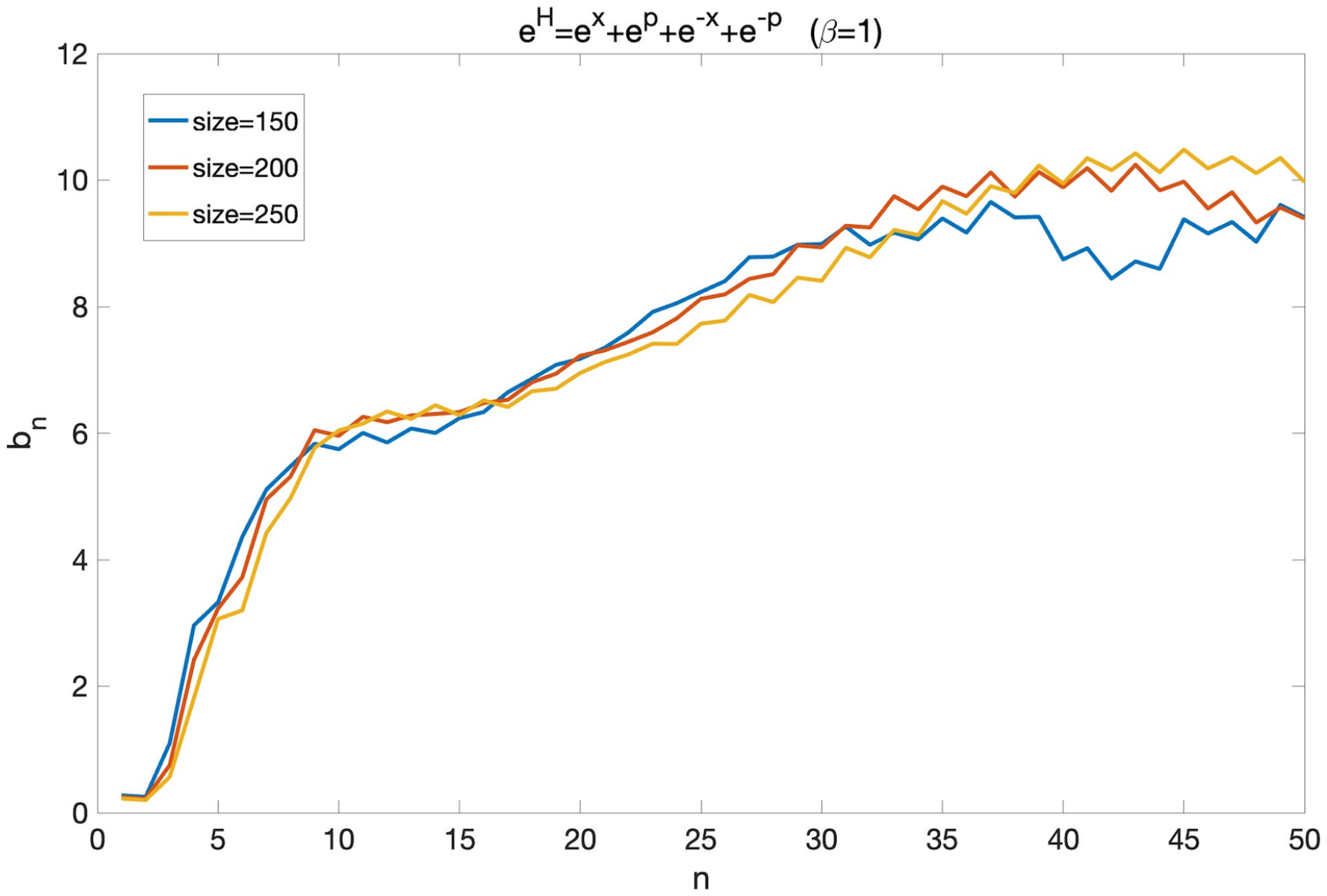}		
	\end{subfigure}

	 \begin{subfigure}{0.49\textwidth}
		\includegraphics[width=\textwidth]{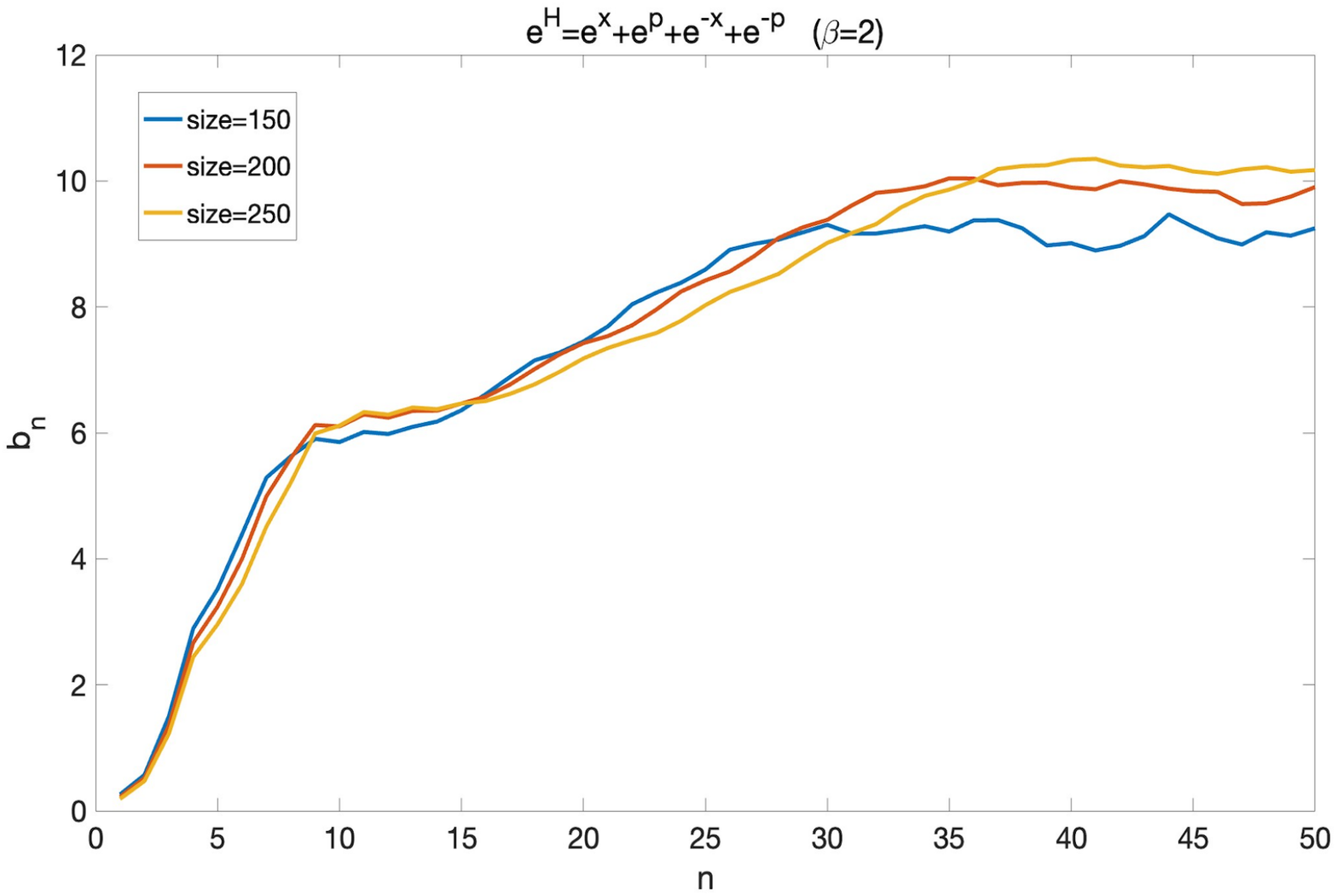} 	
	\end{subfigure}
	\begin{subfigure}{0.49\textwidth}
		\includegraphics[width=\textwidth]{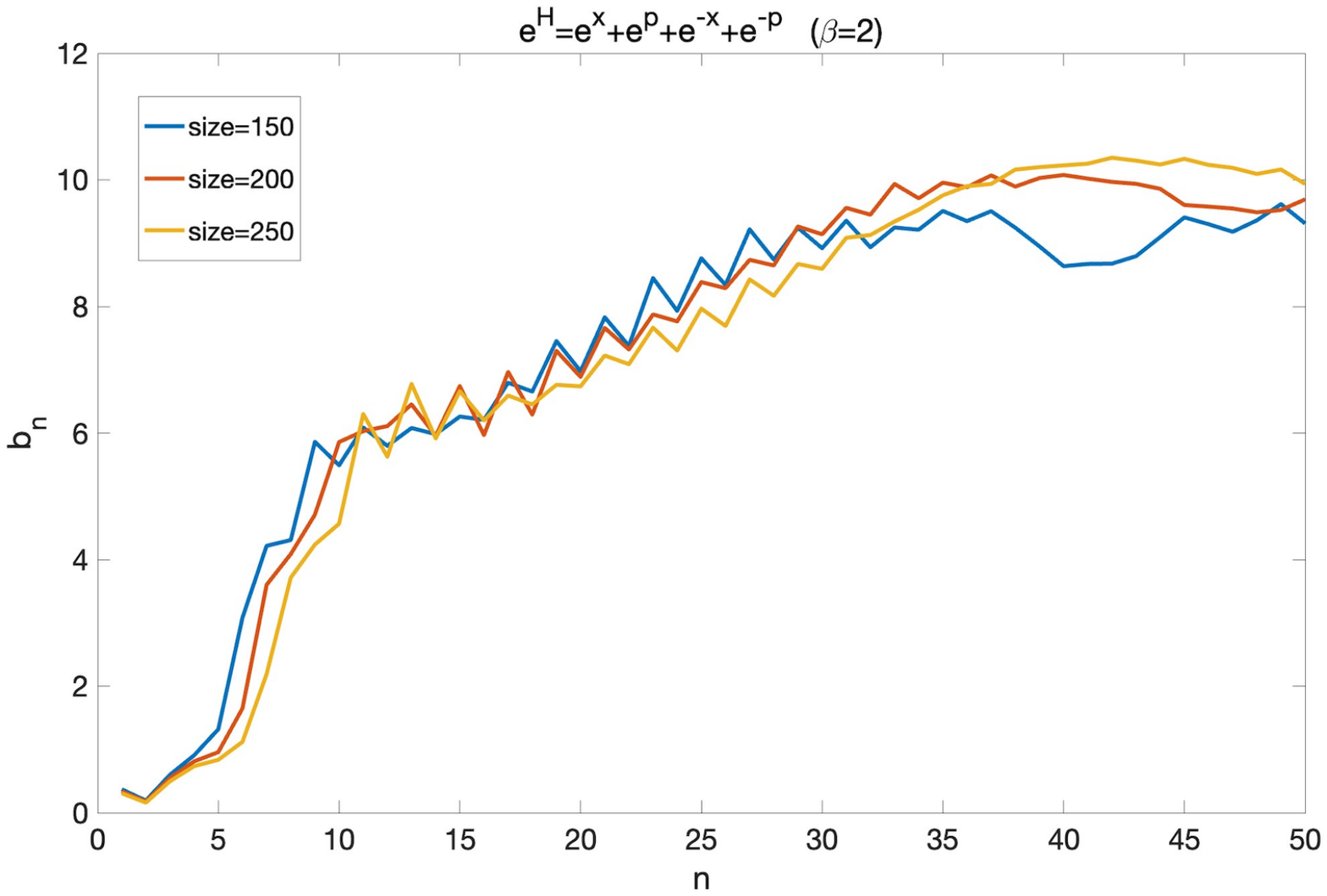}		
	\end{subfigure}

 	\begin{subfigure}{0.49\textwidth}
		\includegraphics[width=\textwidth]{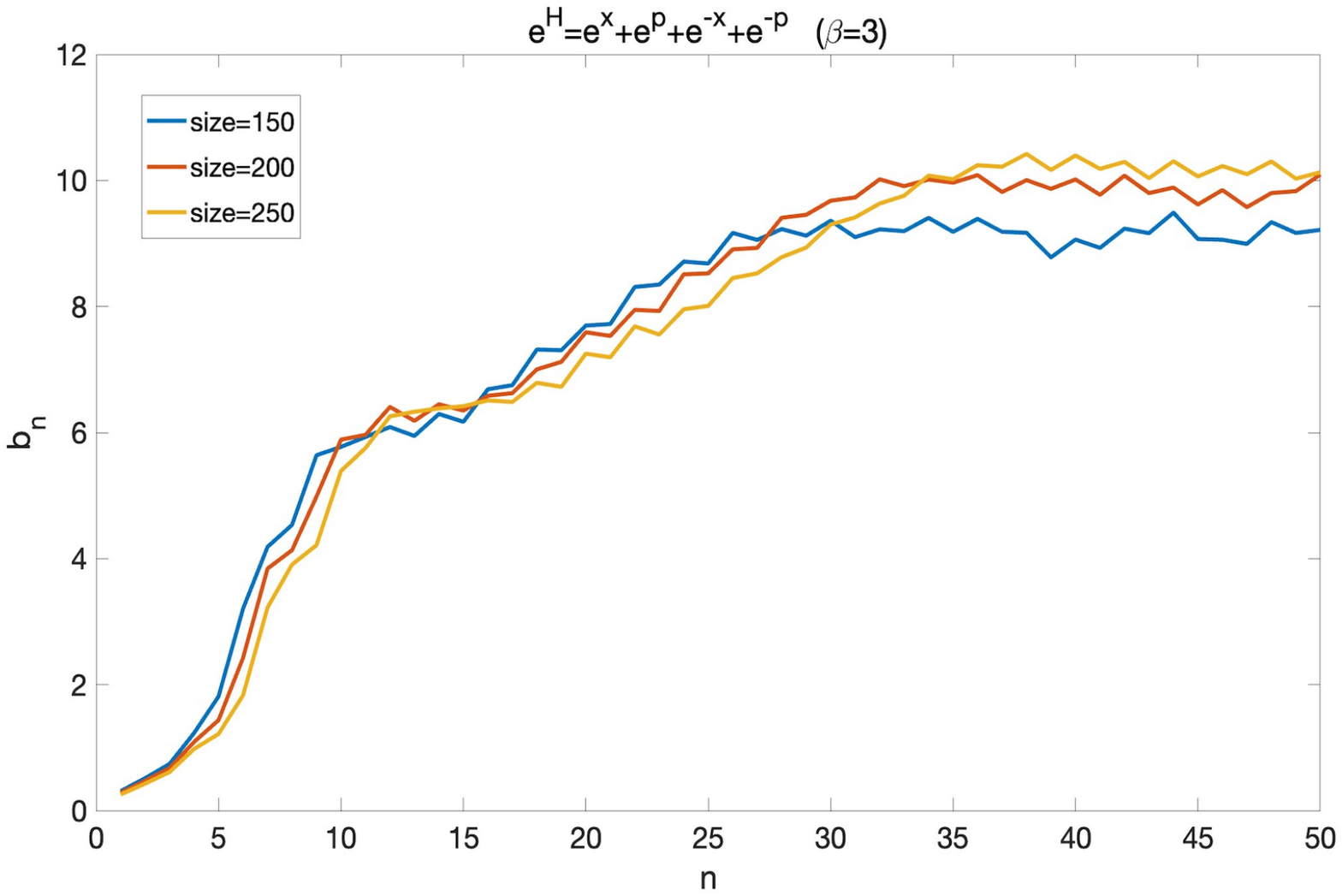} 	
	\end{subfigure}
	\begin{subfigure}{0.49\textwidth}
		\includegraphics[width=\textwidth]{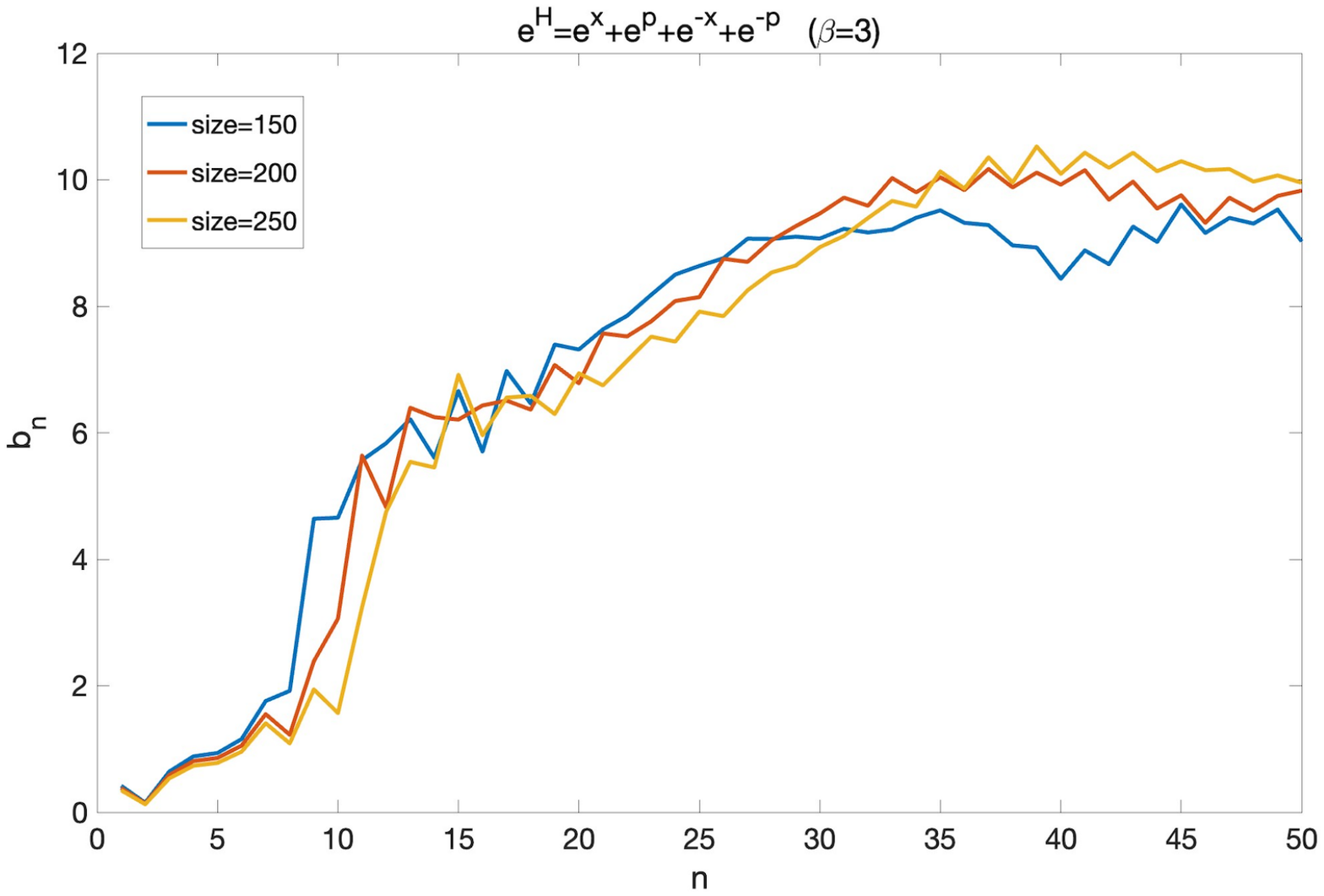}		
	\end{subfigure}
	
 	\begin{subfigure}{0.49\textwidth}
		\includegraphics[width=\textwidth]{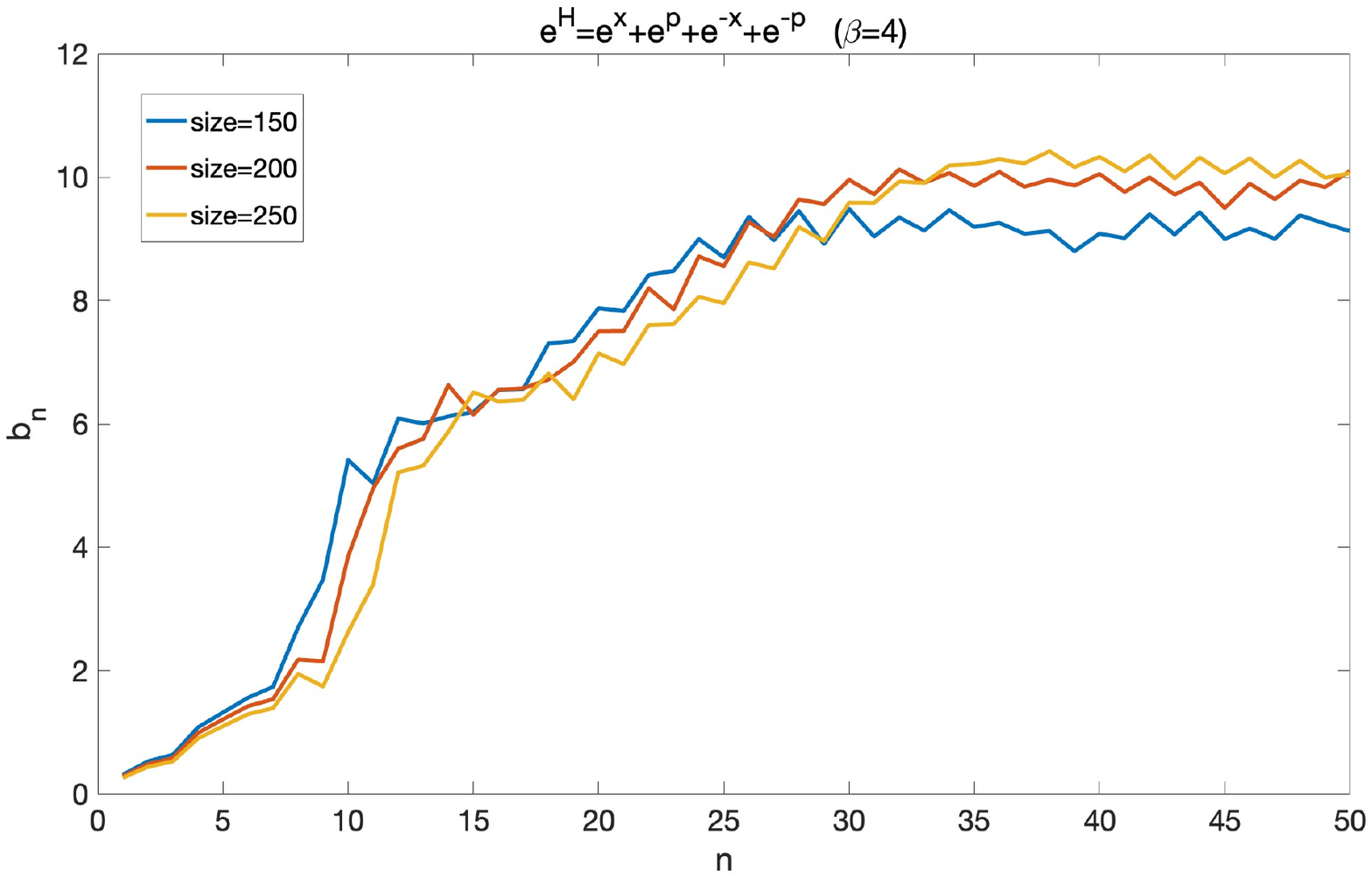} 	
	\end{subfigure}
	\begin{subfigure}{0.49\textwidth}
		\includegraphics[width=\textwidth]{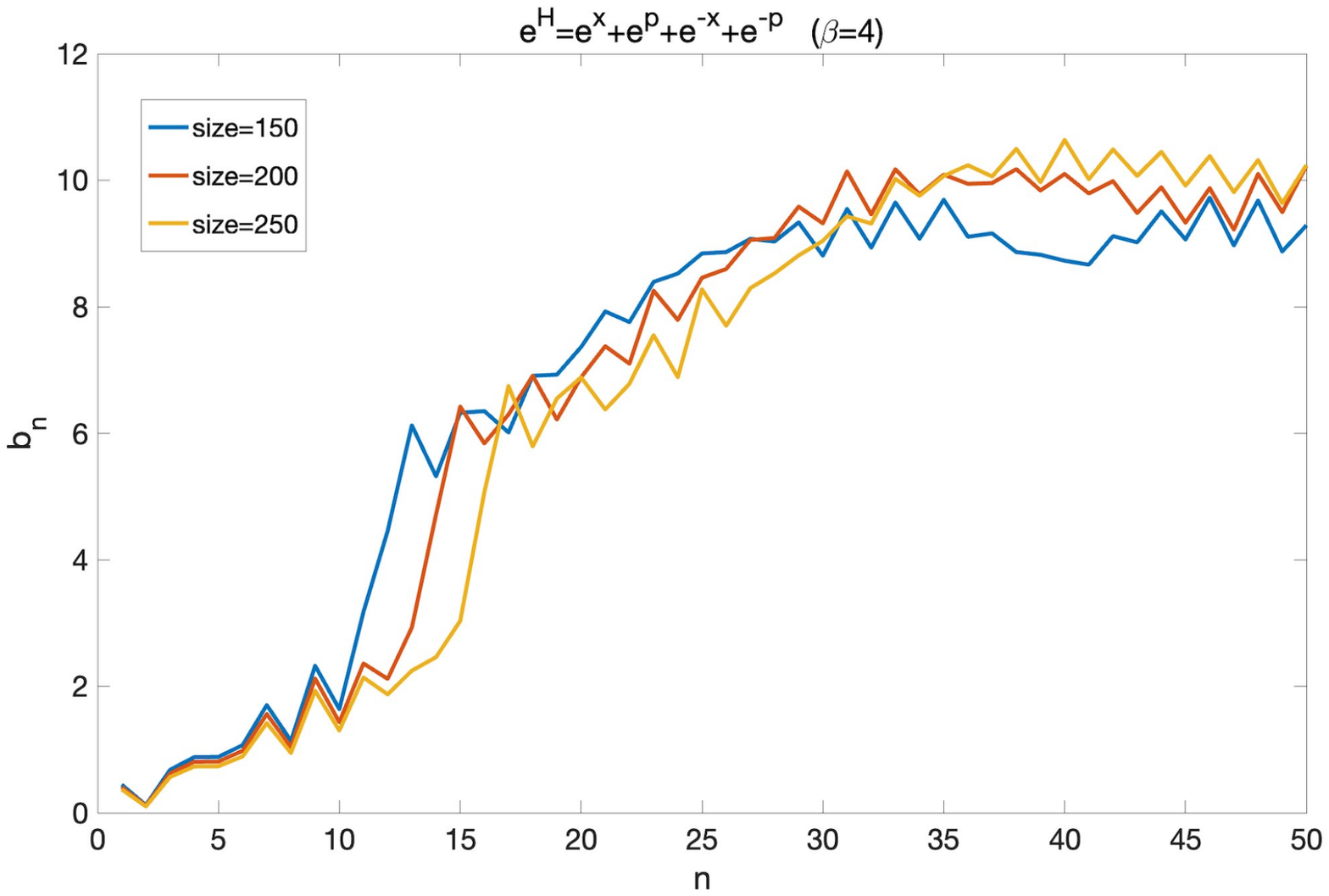}		
	\end{subfigure}

	\caption{The Lanczos coefficients for the local $\mathbb{P}^1\times\mathbb{P}^1$ model. The initial operator for the left figures is $\mathcal{O}_0=e^{\hat{x}}$, and for the right figures is $\mathcal{O}_0=\hat{x}$.}
	\label{figurep1p1}

\end{figure} 

\begin{figure}
	\centering
	\begin{subfigure}{0.49\textwidth}
		\includegraphics[width=\textwidth]{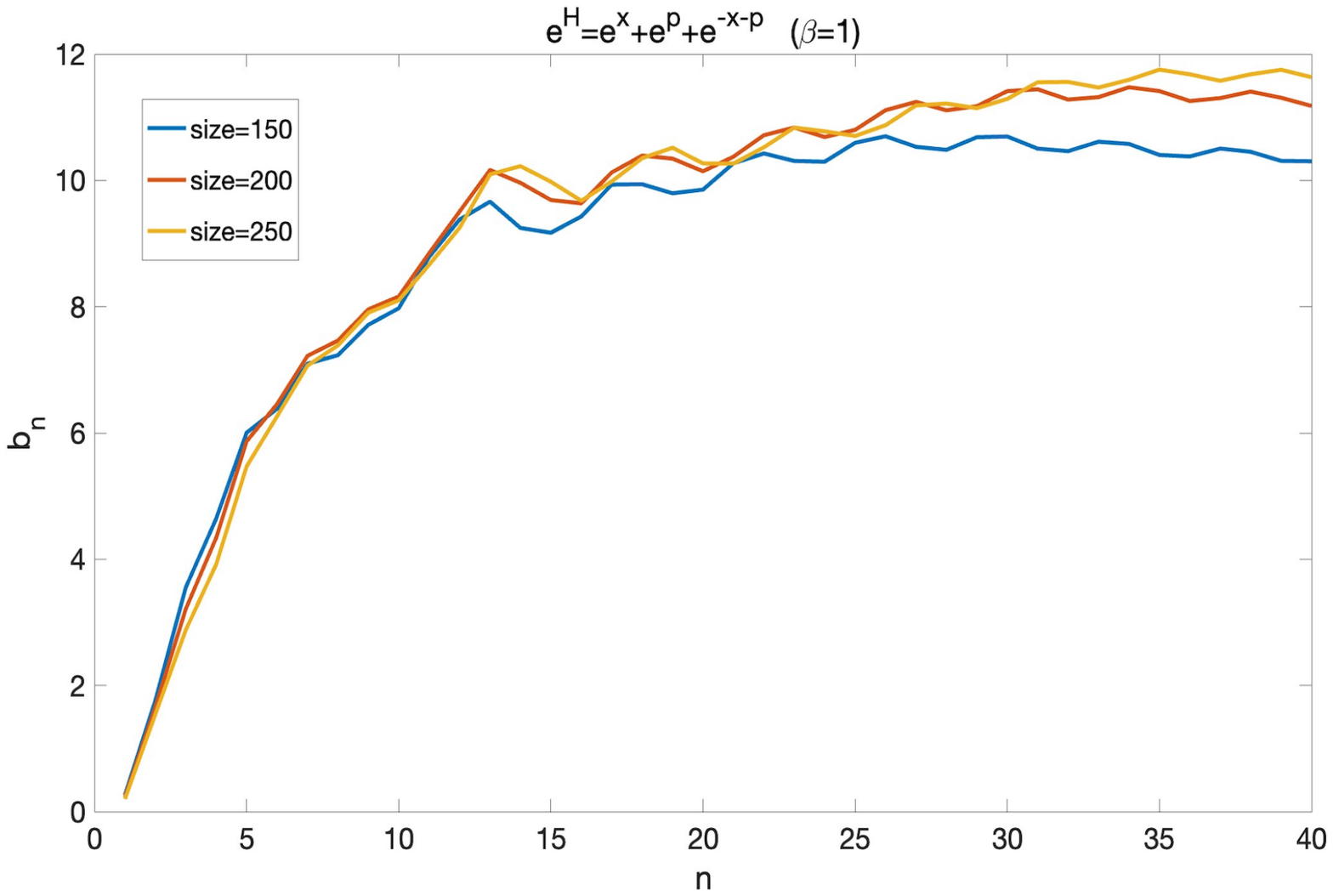} 	
	\end{subfigure}
	\begin{subfigure}{0.49\textwidth}
		\includegraphics[width=\textwidth]{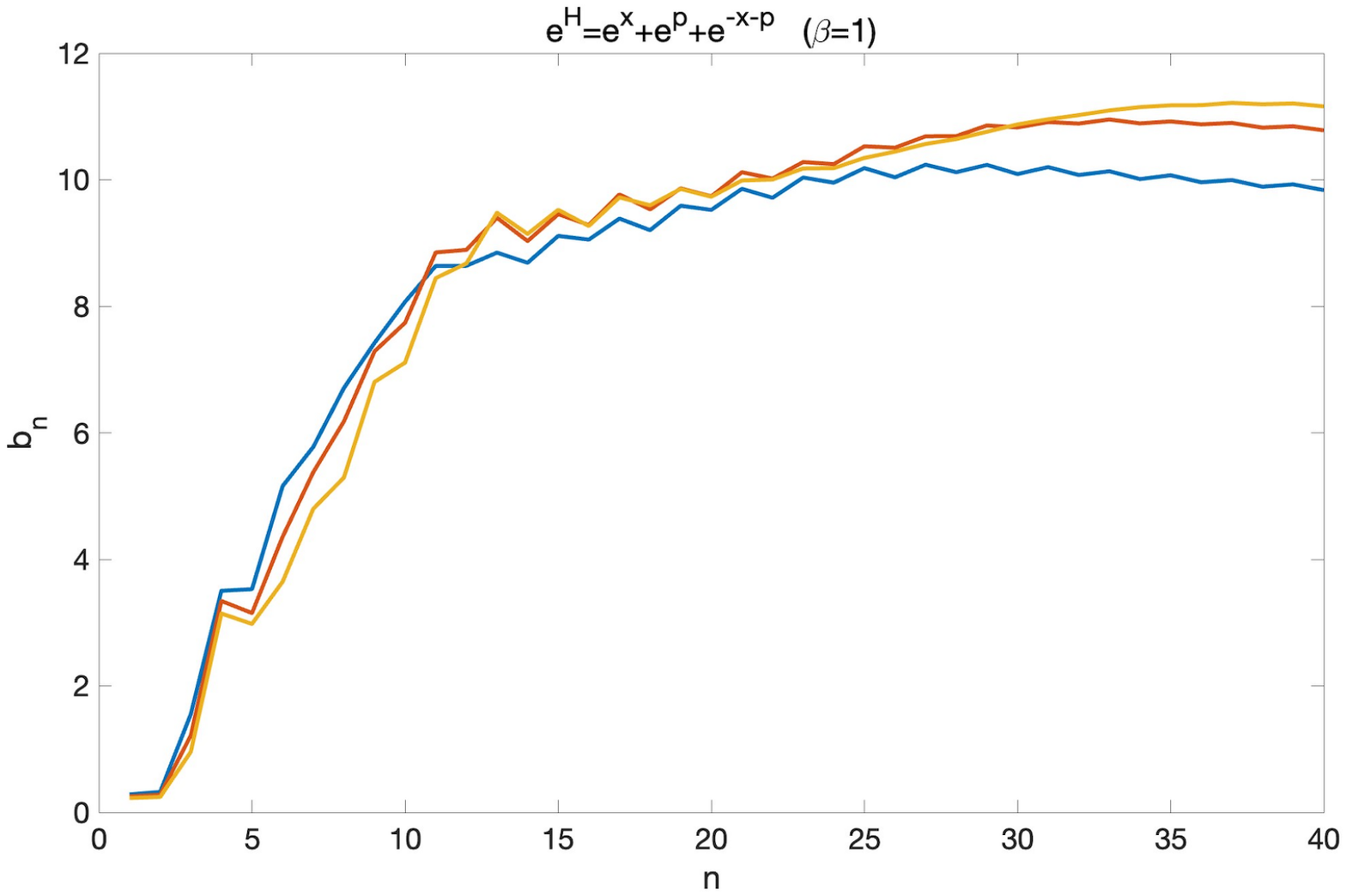}		
	\end{subfigure}

	 \begin{subfigure}{0.49\textwidth}
		\includegraphics[width=\textwidth]{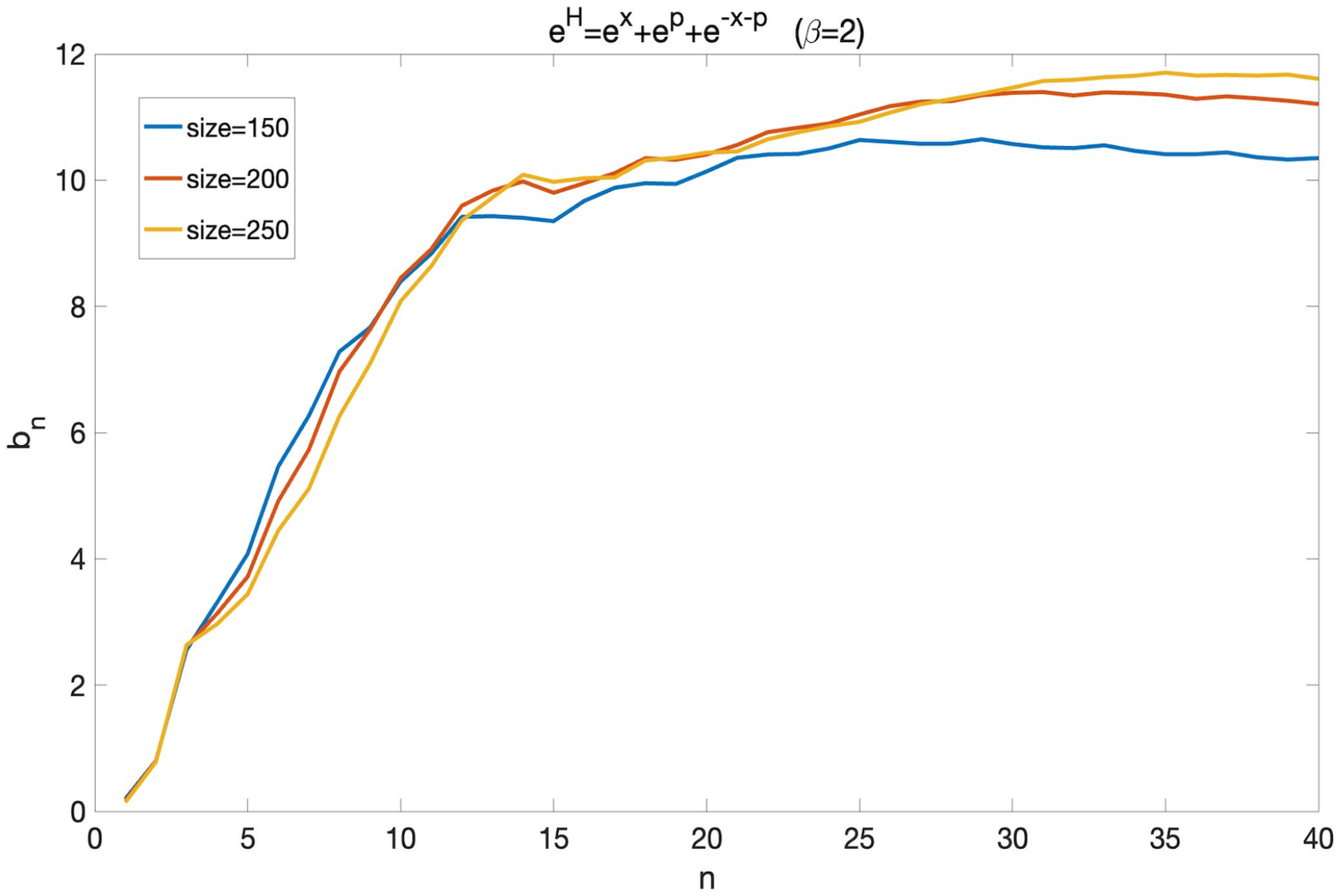} 	
	\end{subfigure}
	\begin{subfigure}{0.49\textwidth}
		\includegraphics[width=\textwidth]{p2_beta_2.pdf}		
	\end{subfigure}

 	\begin{subfigure}{0.49\textwidth}
		\includegraphics[width=\textwidth]{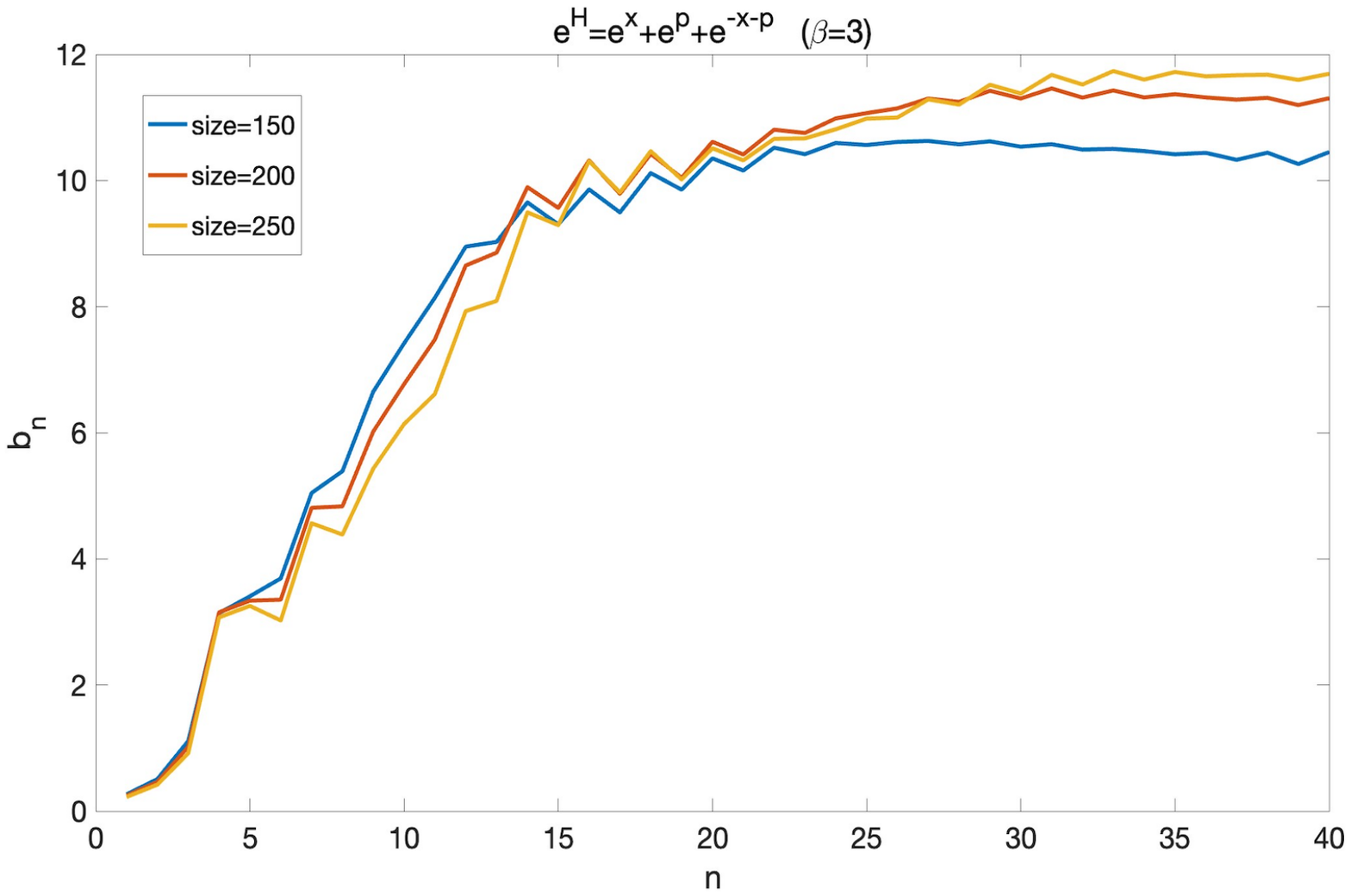} 	
	\end{subfigure}
	\begin{subfigure}{0.49\textwidth}
		\includegraphics[width=\textwidth]{p2_beta_3.pdf}		
	\end{subfigure}

 	\begin{subfigure}{0.49\textwidth}
		\includegraphics[width=\textwidth]{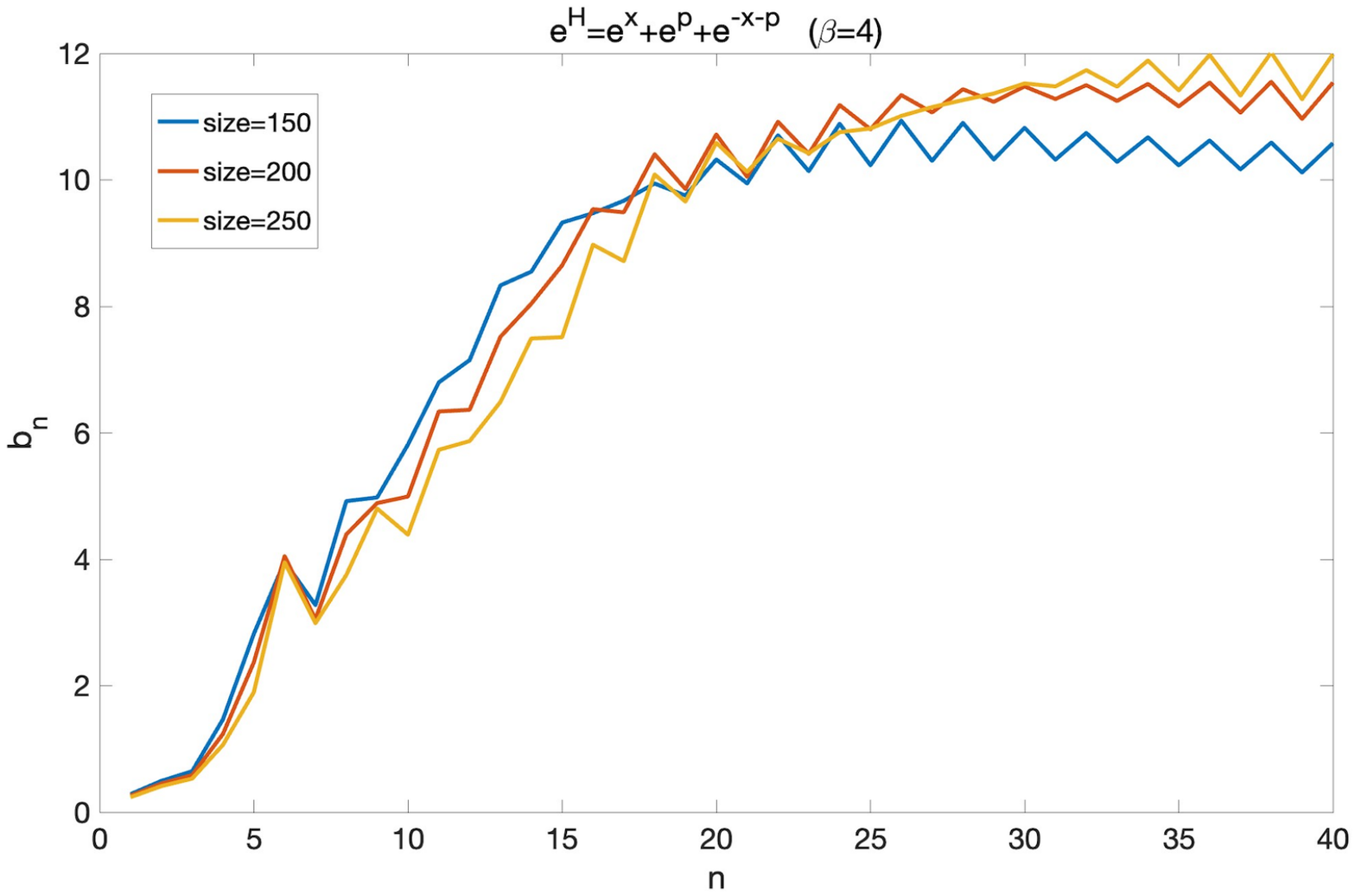} 	
	\end{subfigure}
	\begin{subfigure}{0.49\textwidth}
		\includegraphics[width=\textwidth]{p2_beta_4.pdf}		
	\end{subfigure}

	\caption{The Lanczos coefficients for the local $\mathbb{P}^2$ model. The initial operator for the left figures is $\mathcal{O}_0=e^{\hat{x}}$, and for the right figures is $\mathcal{O}_0=\hat{x}$.}
	\label{figurep2}

\end{figure}

\begin{figure}
	\centering
		\includegraphics[width=0.95\textwidth]{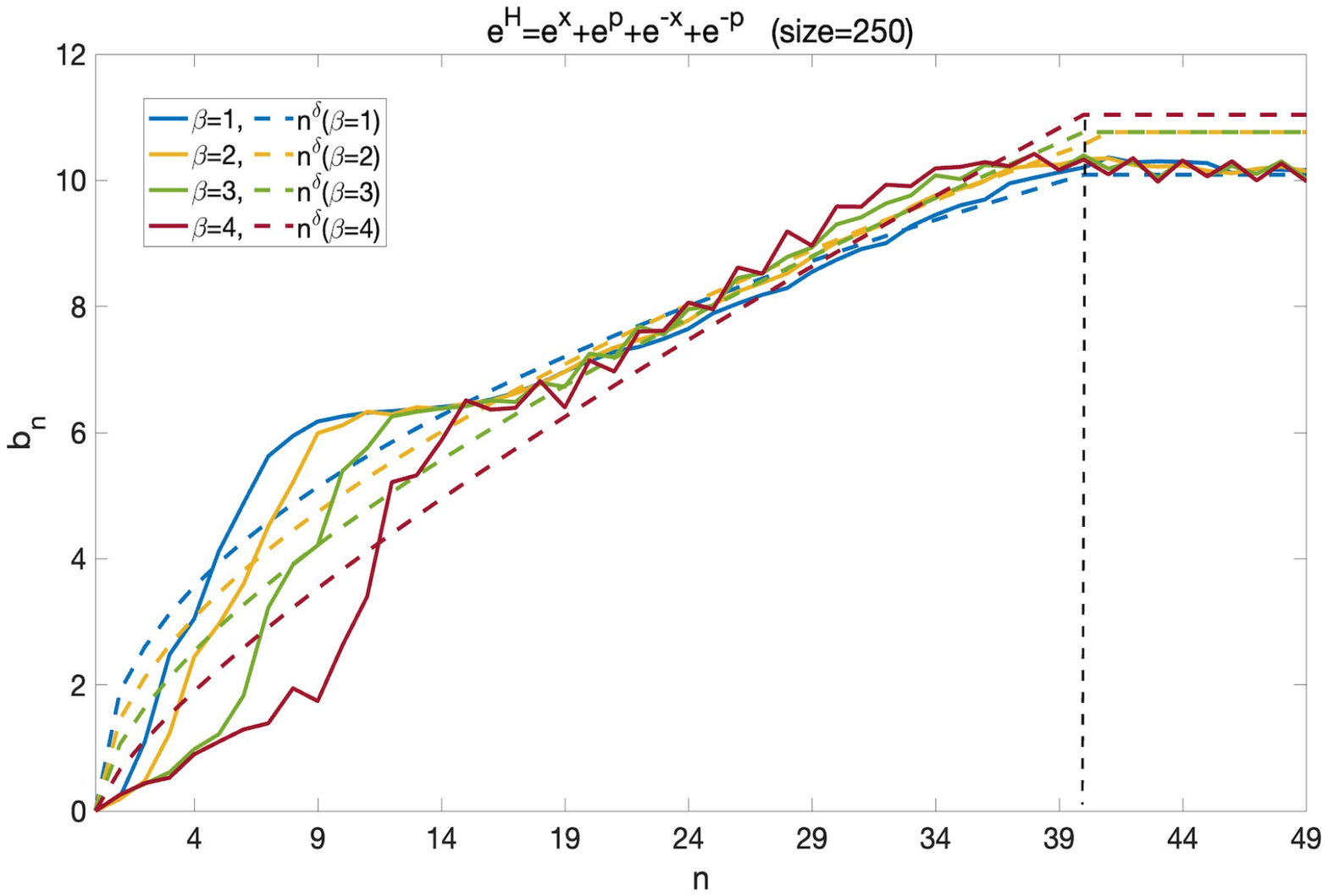} 	
	
		\includegraphics[width=0.95\textwidth]{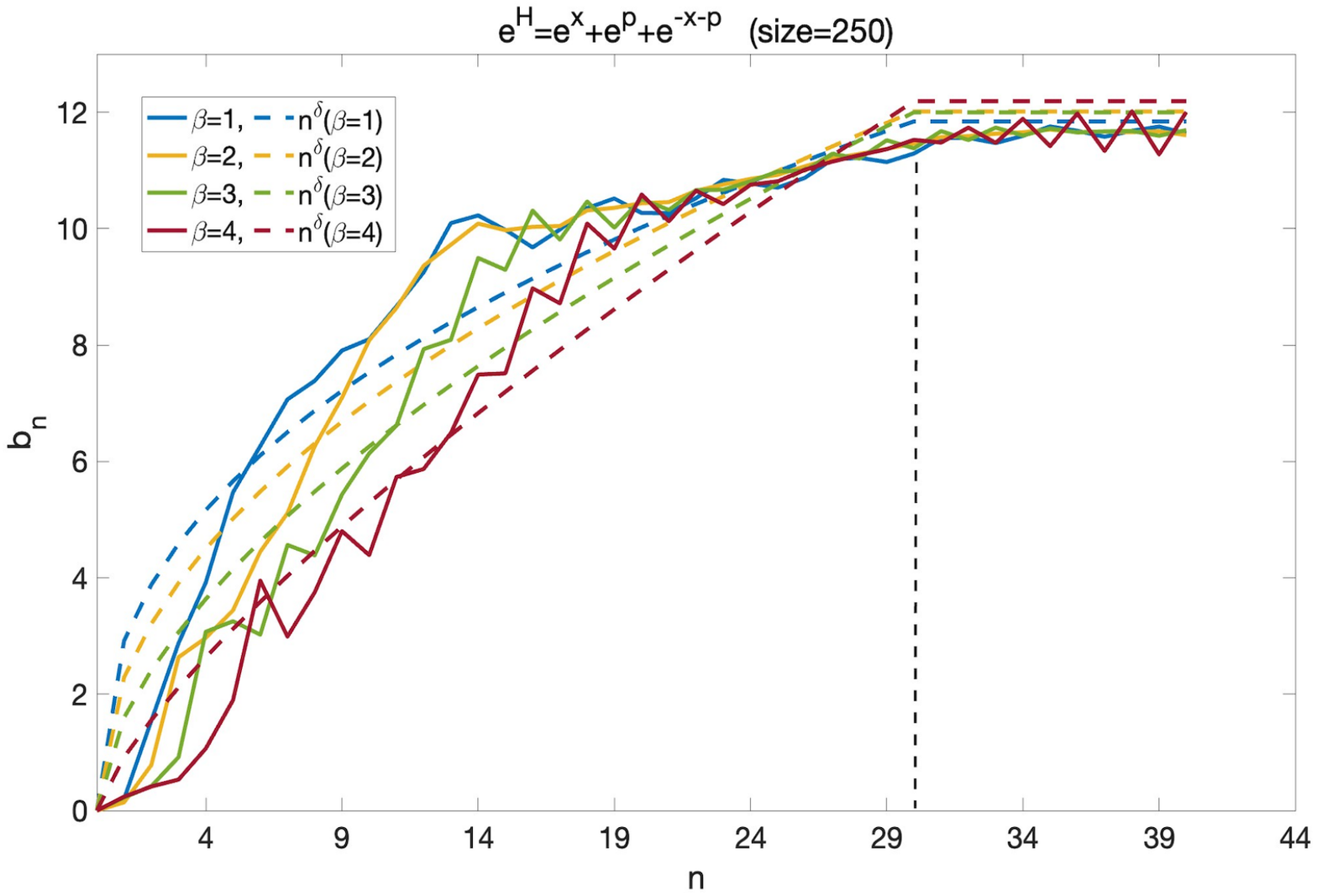}		
\caption{The Lanczos coefficients for the local Calabi-Yau models with the initial operator $\mathcal{O}_0=e^{\hat{x}}$.  We fit the plots with dashed curves of the same color proportional to the power law $n^{\delta}$, in the range of growth before reaching the plateau indicated by a vertical dash line.}
	\label{fit}
\end{figure}

\begin{figure*}
	\centering
			\includegraphics[width=\textwidth]{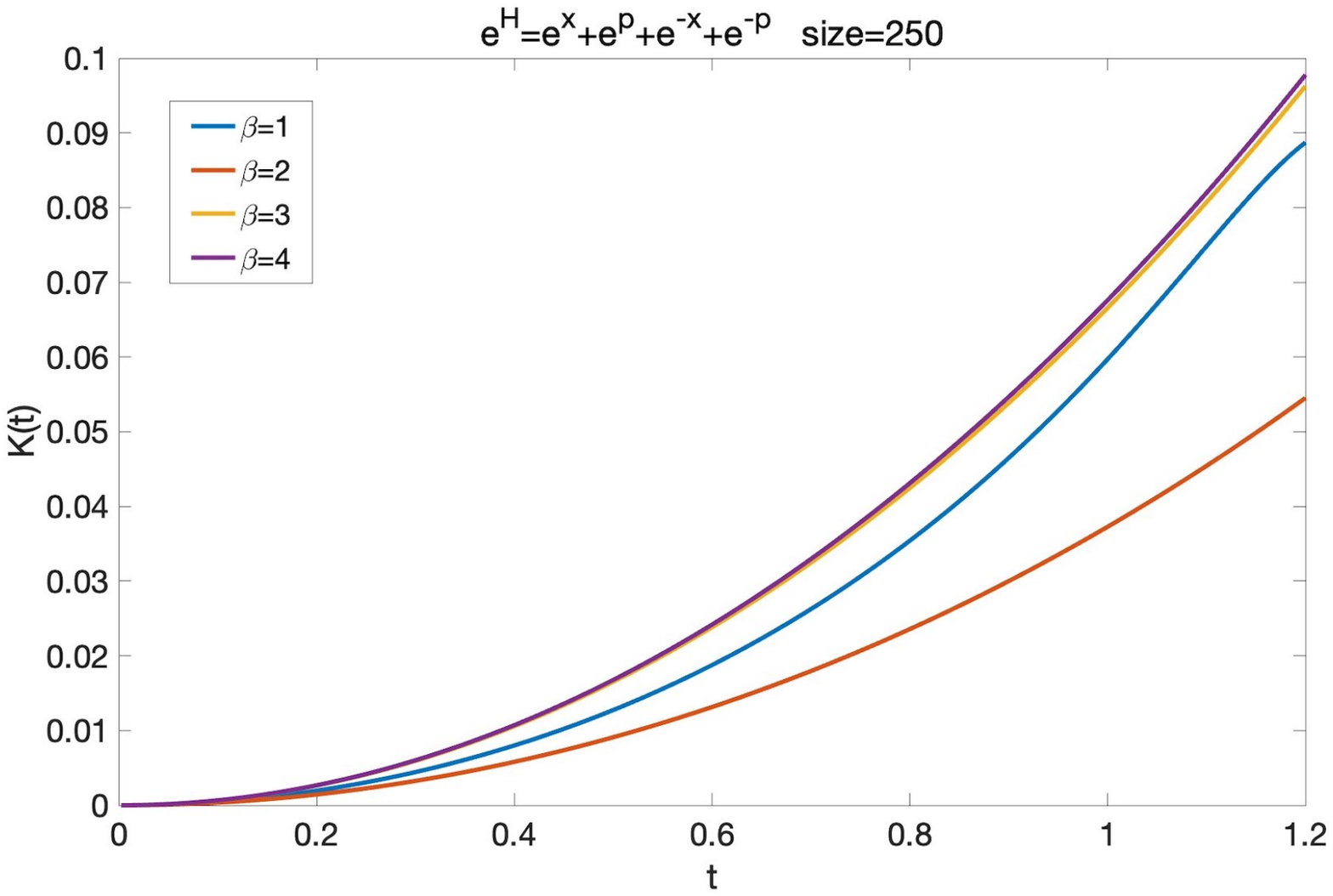} 	

		\includegraphics[width=\textwidth]{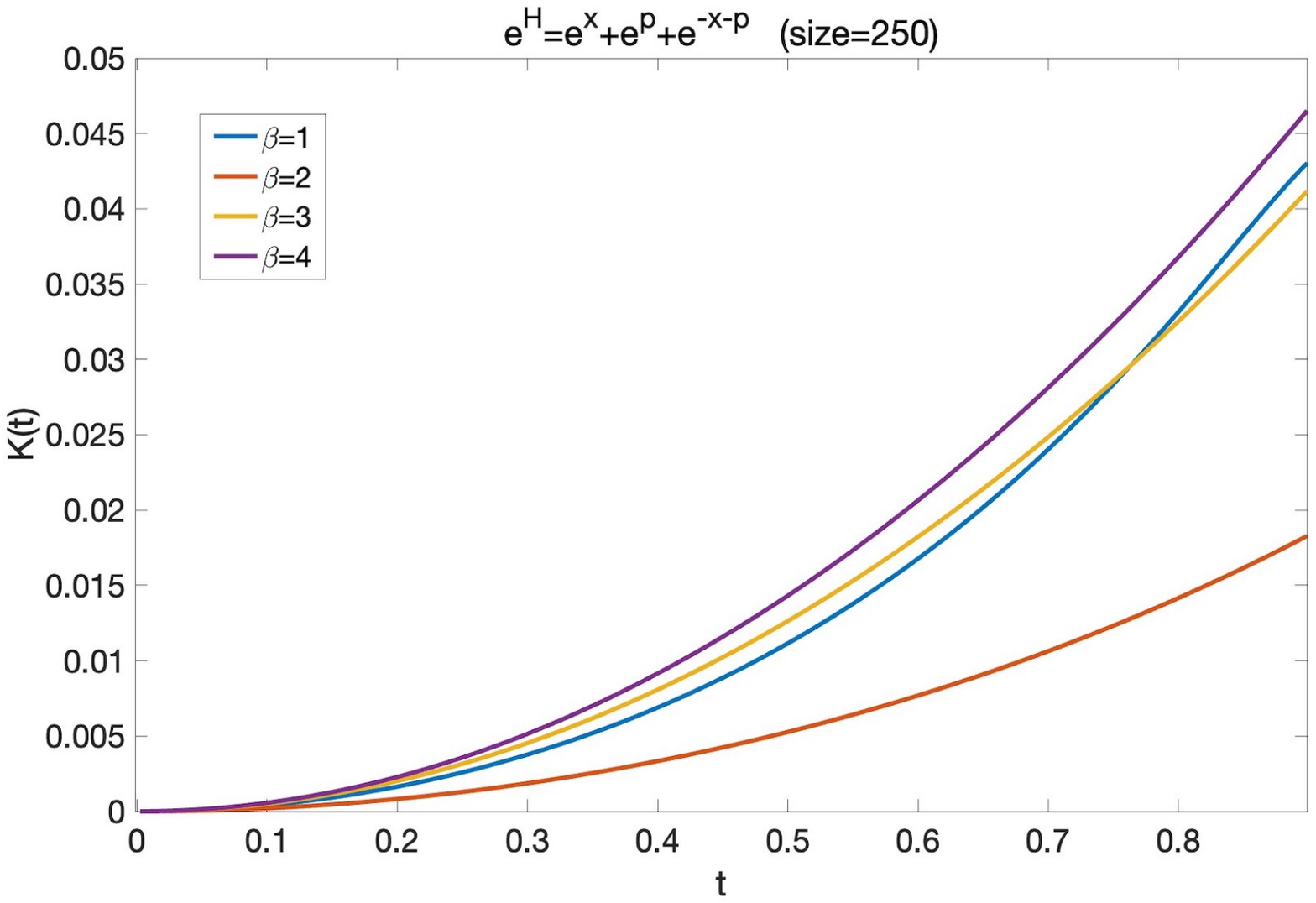} 

	\caption{The Krylov complexity for the Calabi-Yau models}
	\label{complexityCY}
\end{figure*}

Here we will not further explore the analytic approach since it can only possibly work for integer $\beta$ parameters and the integrals still need to be computed numerically. Instead, we use the numerical approach, which is rather straightforward with no restriction on parameters. After solving the energy eigenvalues and eigenstates numerically by truncation in a harmonic oscillator basis, the moments can be computed by 
	\be\ba
		\mu_{2n}&=(\mathcal{O}_0|\mathcal{L}^{n}\mathcal{L}^{n}|\mathcal{O}_0)\\
		&= \frac{1}{\sum_i e^{-\beta E_i}}\sum_{i,j}  e^{-\frac{\beta}{2}(E_i+E_j)} (E_i-E_j)^{2n}|\langle i|\mathcal{O}_0|j\rangle |^2 .
		\label{eq4.4}
	\ea\ee
The calculation of $\mu_{2n}$ is much more straightforward than the norm of $\mathcal{O}_n$ \cite{Guo:2022hui,Baek:2022pkt}. Then the Lanczos coefficients can be solved by the relation (\ref{relation2.11}).

Computing the Krylov complexity is also straightforward. In our calculations, we calculate the Lanczos coefficients $b_n$'s up to a finite order $n\leq L$. Then we expand the autocorrelation function $C(t)=b(it,-it)$ as a power series of $t$ and truncate it at order $2L$. 
Then the  $\phi_n$ functions in (\ref{phi2.7}) can be computed recursively as a polynomial function of $t$ of degree $2L$. It is easy to find that
\be
	\phi_n(t)= \frac{(\prod_{i=1}^n b_i) }{n!} t^n+ c_{n+2}t^{n+2}+\cdots.
\ee
We see that it starts with $t^n$ term with a simple formula for the coefficient and all powers of $t$ have the same parity. 
The normalization relation remains consistent with the truncation of the series expansion 
	\be
		\sum_{i=0}^L |\phi_i(t)|^2 = 1+ O(t^{2L+1}),
	\ee
so we can compute the Krylov complexity from (\ref{Krylov2.10}). This should be a good approximation for the range of small $t$ within the radius of convergence if the truncation order $L$ is large enough.  

We use the truncation method in a basis of harmonic oscillator eigenstates to numerically solve the Calabi-Yau models, we select three different truncation sizes of $150\times150, 200\times200, 250\times250$, and set the parameters $\hbar=1$ and $\beta$ in the range of $1\sim 4$. Note that this truncation size is unrelated to the truncation order $L$ in the calculations of Krylov complexity. In Figure \ref{figurep1p1} and Figure \ref{figurep2}, we choose the initial operator $\mathcal{O}_0=e^{\hat{x}},\hat{x}$, compute the Lanczos coefficients for the two different Calabi-Yau models and compare the results for different truncation sizes in the same graph. One can see that when we increase the truncation size from $150\times150$ to $250\times250$, the shape hardly changes. In fact, the curves are almost visually coincident for small $n$'s, while the differences are more obvious for larger $n$'s after the growth reaches a plateau. So at least for small $n$'s before reaching the plateau, the corrections from finite size effects to our calculations should be negligible for our purpose and the behaviors of the Lanczos coefficients should be credible.  For the two different initial operators, we find their figures of  Lanczos coefficients have  almost the same shape, while the case of  $\mathcal{O}_0=e^{\hat{x}}$ has a slightly smoother behavior. This indicates the shape independence of the initial operators in this case. 

 In Figure \ref{fit}, we plots the Lanczos coefficients of the smoother case $\mathcal{O}_0=e^{\hat{x}}$ with the maximal truncation size $250\times250$. For the different $\beta$ parameters, we see that the plateaus have almost the same height.  We fit the plots with dashed curves of the same color proportional to the power law $n^{\delta}$, in the range of growth before reaching the plateau indicated by a vertical dash line.  For the local $\mathbb{P}^1\times\mathbb{P}^1$ model, the Lanczos coefficients reach the plateau at around $n=40$, while for the local $\mathbb{P}^2$ model, it is around $n=30$. The growth of Lanczos coefficients in the local $\mathbb{P}^2$ model appears to be smoother than the local $\mathbb{P}^1\times\mathbb{P}^1$ model, and fits better with the power law $n^{\delta}$.  The best fit results for the $\delta$ parameter are listed in Table \ref{delta1}.  Of course, the fits are not visually quite perfect,  but it should be clear from the pictures that the growth is not linear, especially in comparisons with the models in the subsequent Sections \ref{quartic},  \ref{Toda}. Furthermore, the slopes of a hypothetical and seemingly worse linear fit to the curves in Figure \ref{fit} would not be proportional to the temperatures, unlike the non-relativistic models in the subsequent sections. For integrable systems, the Lanczos coefficients should indeed grow slower than linearly before reaching the plateau. But as we mentioned in the introduction Section \ref{sectionintro}, it is also known that the saddle-dominated scrambling can produce a seemingly linear growth in integrable systems \cite{Xu:2019lhc,Bhattacharyya:2020art}.  The plausible explanation here seems to be that the scrambling effect contributed by saddles is minor for the Calabi-Yau models but much dominant for the models in subsequent Sections \ref{quartic},  \ref{Toda}. It would be certainly interesting to have a better understanding of the issue in future research.   Examples of non-linear growth of Lanczos coefficients have appeared in the literature, e.g. recently in the study of the Lipkin-Meshkov-Glick model in critical quench scenarios \cite{Afrasiar:2022efk}.

We should note that there are some obvious fluctuations of the Lanczos coefficients, which is larger for the initial operator $\mathcal{O}_0=\hat{x}$. The fluctuations will be also more evident in the quartic anharmonic oscillator model in the next Section \ref{quartic}, but is less evident in the non-relativistic Toda model in Section \ref{Toda}. Such oscillations of $b_n$'s are seen to usually depend on the parity of $n$, and have been explained in  \cite{Yates:2020lin,Yates:2020hhj,Bhattacharjee:2022vlt, Bhattacharjee:2022ave, Baek:2022pkt} as due to a small contribution from the autocorrelation function with the sign $(-1)^n$. However, we should note that in our models, the oscillation is not always very regular and the period is not always 2. The oscillation is also sensitive to the truncation size in our numerical calculations. It would be interesting to have a more precise understanding of the phenomenon.

In the Figure \ref{complexityCY}, we plot the Krylov complexity for the Calabi-Yau models for different $\beta$ parameters. An intriguing feature is that the arrangement of the curves may not necessarily be in the order of $\beta$ parameters. Essentially, for small $t$ the Krylov complexity depends most on Lanczos coefficients $b_n$'s of small $n$'s.  For the case of $\beta=2$, the Lanczos coefficient $b_1$ is smaller than others when the truncation size is increased to 250, so the curve is lowest. However the feature depends quite sensitively on the truncation size of the numerical calculations, and it would be interesting to have a better understanding. 

As we mention earlier that the growth behavior of the Krylov complexity should be $K(t)\sim t^{\frac{1}{1-\delta}}$. However, it is also known that the time dependence of Krylov complexity at very short time is not a reliable way to extract the parameter $\delta$, which should be computed more directly from the Lanczos coefficients. \footnote{This problem appears in the first version of the paper. Using the correct method, the results for the parameter $\delta$ are now modified.}

\begin{table}
\begin{center} 
	\begin{tabular} 
	{|c|c|c|c|c|} \hline  Calabi-Yau models  & $\beta=1$ & $\beta=2$ & $ \beta=3 $ & $\beta=4$ \\
		 \hline  $\mathbb{P}^1\times\mathbb{P}^1$    & 0.4525  & 0.5373 &0.6071 & 0.6583 \\
		  \hline  $\mathbb{P}^2$       &  0.3486  & 0.4007 & 0.5301 &0.5837\\
		 \hline
	\end{tabular}
	\caption{  The best fit values of $\delta$ for the Calabi-Yau models with the initial operator $\mathcal{O}_0=e^{\hat{x}}$ and truncation size $250\times250$} 
	\label{delta1}
\end{center}
\end{table}

\section{The quartic anharmonic oscillator}
\label{quartic}

In this section, we consider the quartic anharmonic oscillator, the Hamiltonian is given by
	\be
		H=p^2+x^2+ gx^4 .
	\ee
The case of a negative quadratic term gives a double-well potential and is consider recently in \cite{Baek:2022pkt}. This well known model has been studied in a long history of literature, see e.g. the early work \cite{Bender:1969si}. 

There is an extra parameter $g$ comparing to the Calabi-Yau models in the previous section. In the Calabi-Yau models, the Planck constant $\hbar$ measures the strength of quantum effects and we have made a specific choice $\hbar=1$. In the current case, we can rescale the position and momentum operators to set $\hbar=1$, and the strength of quantum effect can be absorbed into the coupling parameter $g$, while the temperature parameter $\beta$ also scales accordingly when we perform the Lanczos algorithm. We will perform the numerical calculations with $\hbar=1$ and $g$ in the range of $\frac{1}{20}$ and $\frac{1}{50}$.

\begin{figure*}
	\centering
	
			\includegraphics[width=\textwidth]{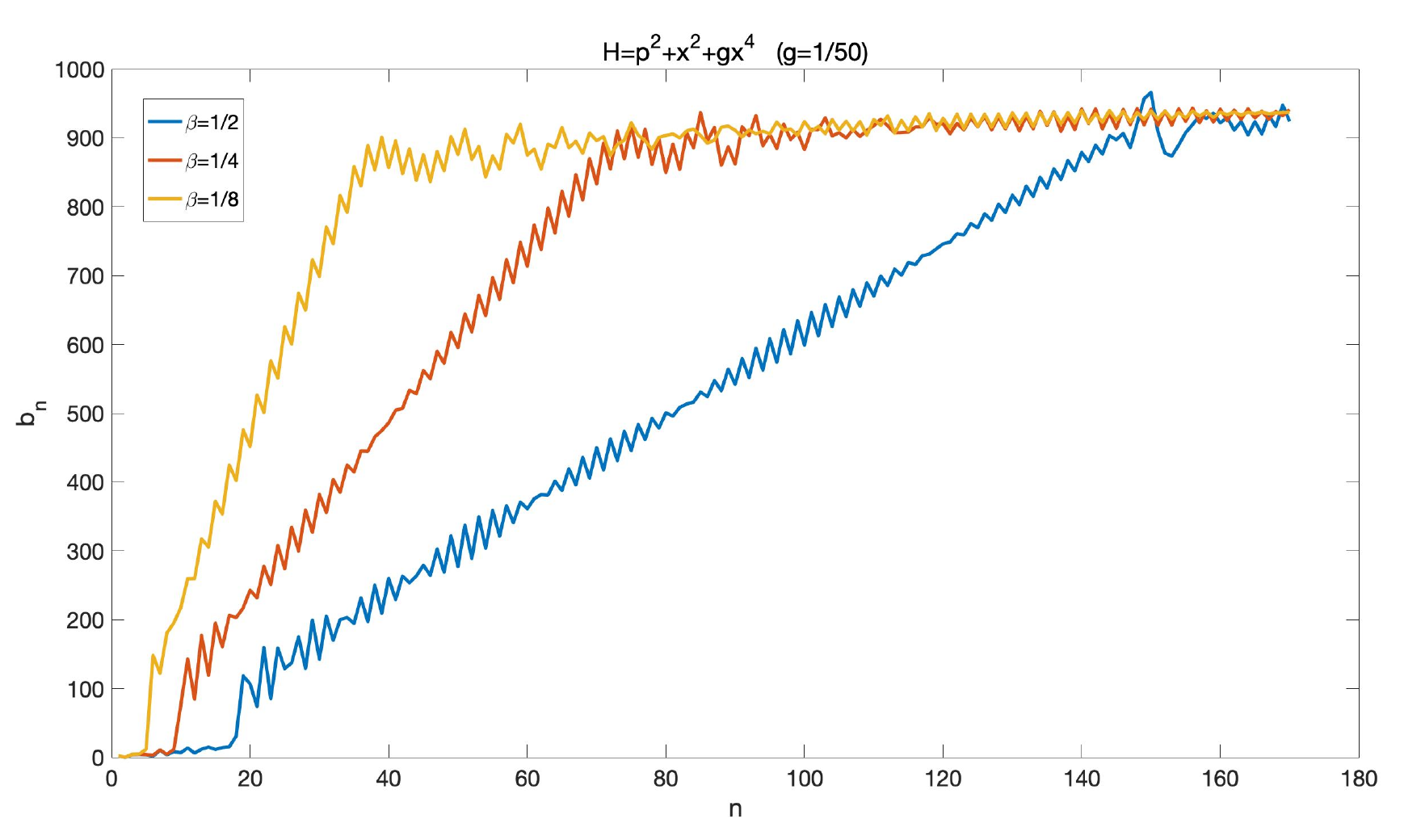} 	

	 \vskip 18pt

		\includegraphics[width=\textwidth]{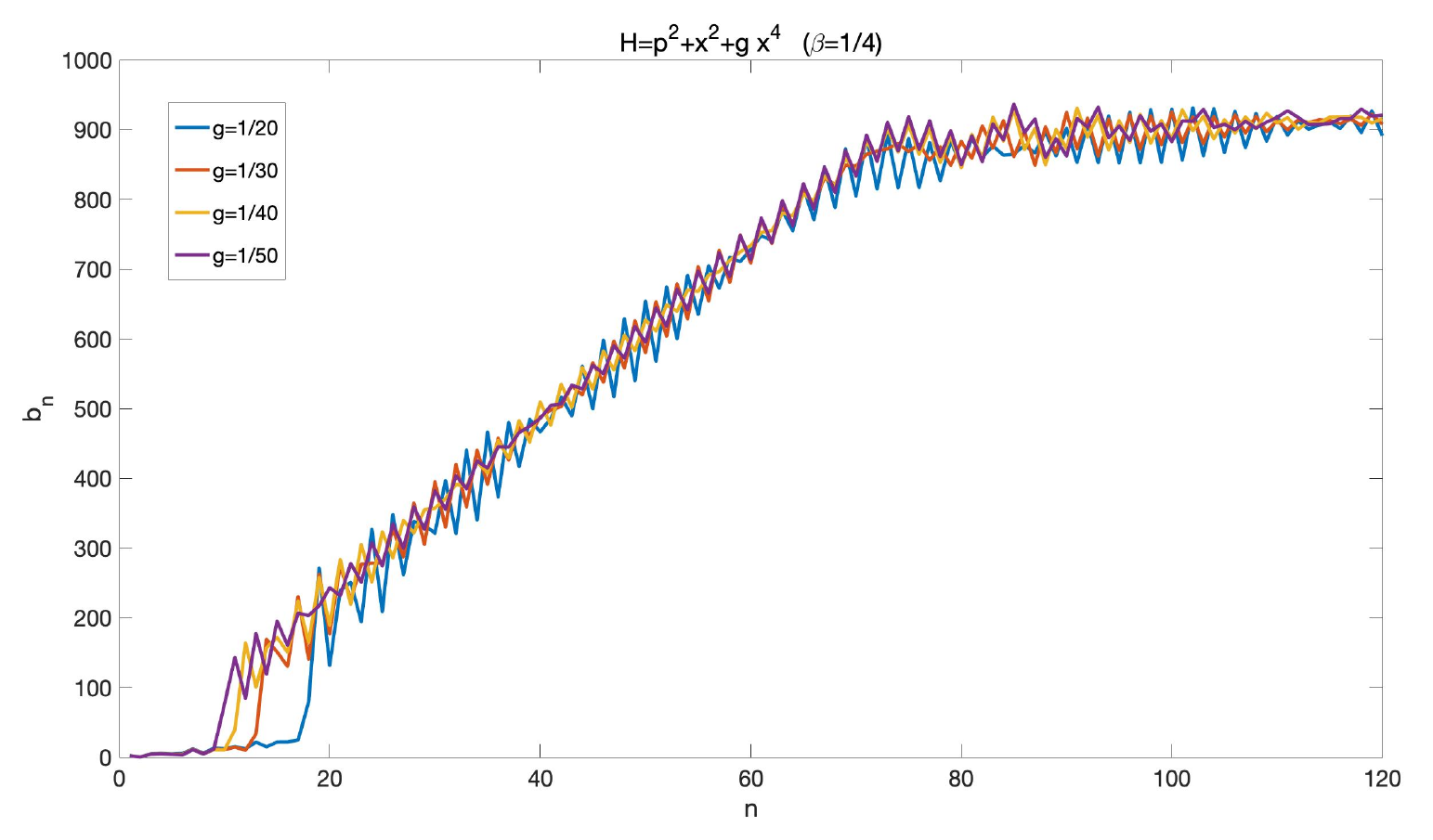} 

	\caption{The Lanczos coefficients for the quartic anharmonic oscillator model}
	\label{x4diagram}

\end{figure*}

We choose the initial operator $\mathcal{O}_0=\hat{x}$ and calculate the Lanczos coefficients with increasing truncation sizes 150, 200, 250. We find that they have almost the same shape before reaching the plateau, similar to the Calabi-Yau models. In Figure \ref{x4diagram}, we give some results about the Lanczos coefficients with the maximal truncation size 250 from the numerical method. We see that the Lanczos coefficients grow linearly,  and finally reach the plateau. The linear growth may come from saddle-dominated scrambling  and does not necessarily indicate that the model is chaotic \cite{Xu:2019lhc,Bhattacharyya:2020art}. The slopes in the linearly growing phase is proportional to $T=\beta^{-1}$ in the range of parameters that we calculate.  Furthermore, in the first figure for a fixed $g=\frac{1}{50}$, we list some results of ratios of the slope with temperature $\alpha \beta$ in Table \ref{bnx4}, and find that they are less than but very close to $\pi$, almost saturating the bound (\ref{bound3.3}). The same phenomenon also appears in the inverted harmonic oscillator in \cite{Baek:2022pkt}, and as mentioned there, this is very different from the truly chaotic model such as the large charge Sachdev-Ye-Kitaev model studied in \cite{Parker:2018yvk}. 

The system with saddle-dominated  scrambling shows the chaotic nature inside an integrable system. The paper \cite{Bhattacharjee:2022qjw} also study an interesting opposite phenomenon that where the integrable nature can lie inside a chaotic system, by showing the scar states hosted inside a chaotic system characterized by the Krylov complexity.

In the second figure, we fix $\beta=\frac{1}{4}$ and vary the $g$ parameter. It shows that the Lanczos coefficients are quite insensitive to $g$ at the range of $\frac{1}{20}\sim\frac{1}{50}$. 

\begin{figure*}
	\centering
			\includegraphics[width=\textwidth]{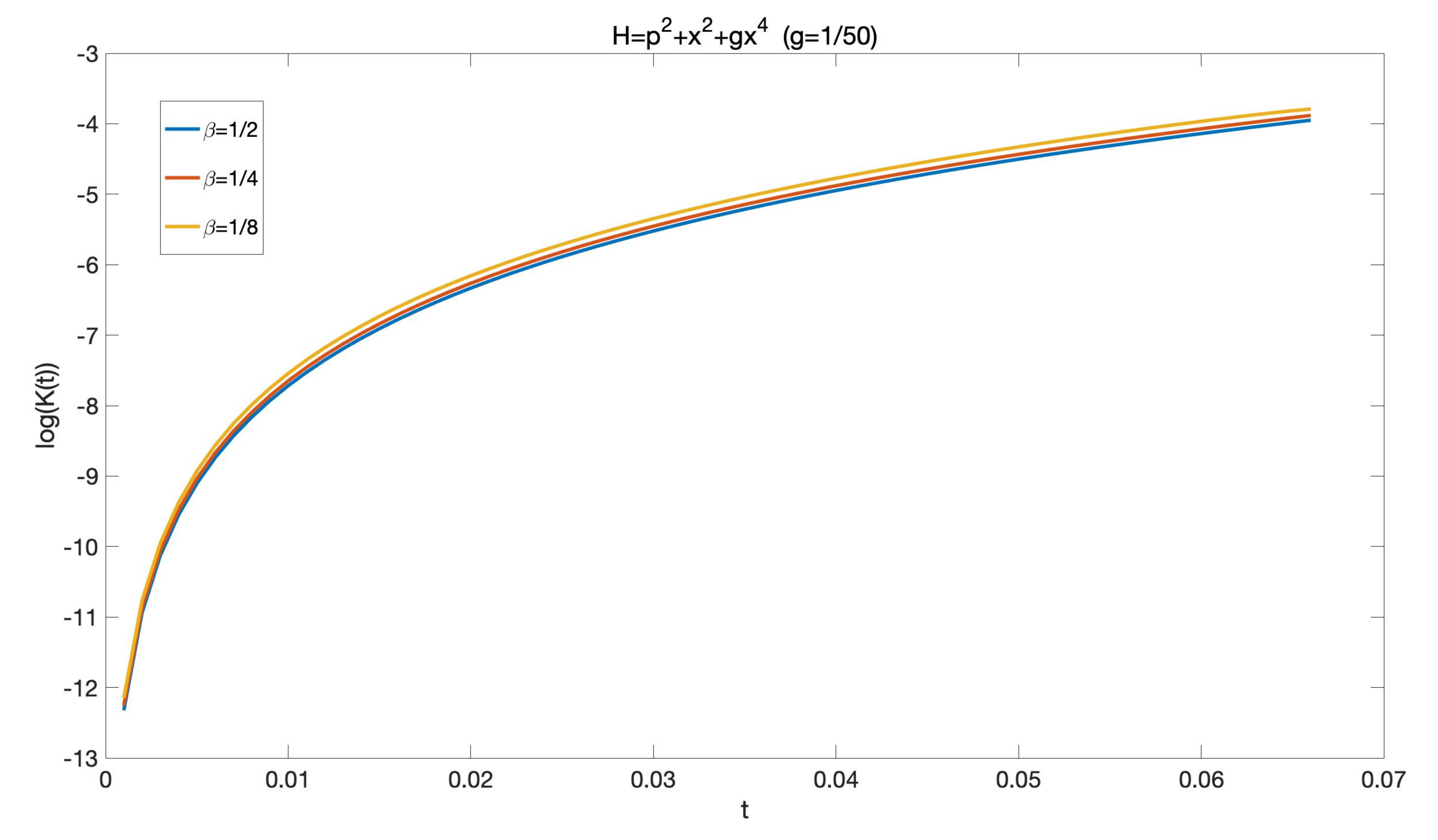} 	

	\caption{The Krylov complexity for the quartic anharmonic oscillator model}
	\label{complexityx4}

\end{figure*}

\begin{table}
\begin{center}

	\begin{tabular} {|c|c|c|c|} \hline  $\beta$   & $\frac{1}{2}$ & $\frac{1}{4}$ & $\frac{1}{8}$ \\
		 \hline  $\alpha$     & 6.1613  & 12.1781 & 23.7828 \\
		  \hline  $\alpha \beta$  &  3.0807  & 3.0445 & 2.9729\\
		 \hline
	\end{tabular}
	
	\caption{  The slope $\alpha$ for the quartic anharmonic oscillator model}
	\label{bnx4}
\end{center}

\end{table}

In Figure \ref{complexityx4}, we plot the $\log(K(t))$ for different $\beta$'s. For different $g$'s, Lanczos coefficients are almost the same in the range of linear growth, so the curves of $\log(K(t))$ also almost coincide with each other. We plot only the case of $g=\frac{1}{50}$ with $\beta=\frac{1}{2},\frac{1}{4},\frac{1}{8}$. Like the paper \cite{Baek:2022pkt}, in our case, the exponential growth is not clearly visible either.

It is an appropriate place here to discuss some technical details of the numerical calculations. Generally, the Lanczos coefficients $b_n$ is very difficult to solve analytically and the recursive method will have large error for as $n$  increases. We solve the Lanczos coefficients from the moments  $\mu_{2n}$'s, which are computed by the equation (\ref{eq4.4}). It may seem that for large energy $E_i$, the contribution is suppressed by an exponential factor so the truncation size does not need to be too large. However if $n$ is large,  the power factor $(E_i-E_j)^{2n}$ may be also very large. So the realistic truncation size around 200 with our reasonable computational resource may also produce a large numerical error in this case. Furthermore, as also encountered in our previous work \cite{Du:2021hfw}, the frequency of the harmonic oscillator used for the truncation calculations need to be empirically appropriately chosen, so that the calculations may converge faster when we increase the truncation size.  

In the previous section, we have noticed that as $\beta$ increases, the region of growth before reaching plateau is larger, which means  we need to compute more $b_n$'s. But when $n$ is large, the calculations of $b_n$ will become more inaccurate as we explained in the above paragraph. On the other hand, if $\beta$ is small, the larger energy eigenvalues will contribute to the Wightman inner product, so we have to increase the truncation matrix size in the basis of eigenfunctions of harmonic oscillator. We choose the parameters appropriately to balance these competing effects so that we may have more accurate numerical results with the range of Lanczos coefficients reaching into the plateau.


\section{A non-relativistic Toda model}
\label{Toda}

In this section, we consider another class of integrable models, the non-relativistic Toda models. In particular, we focus on the simplest two-body case of the Toda chain models, where the center of motion has trivial dynamics and does not appear in the Hamiltonian. The Hamiltonian for the relative two-body motion is 
	\be
		\hat{H}=\frac{\hat{p}^2}{2}+2g\cosh (\hat{x}).
	\ee
The exact quantization condition was first derived in \cite{Gutzwiller:1980yx}. In the modern approach it is related to the pure $SU(2)$ Seiberg-Witten gauge theory \cite{Nekrasov:2009rc, Grassi:2019coc}.

\begin{figure*}
	\centering
	\begin{subfigure}{1\textwidth}
		\includegraphics[width=\textwidth]{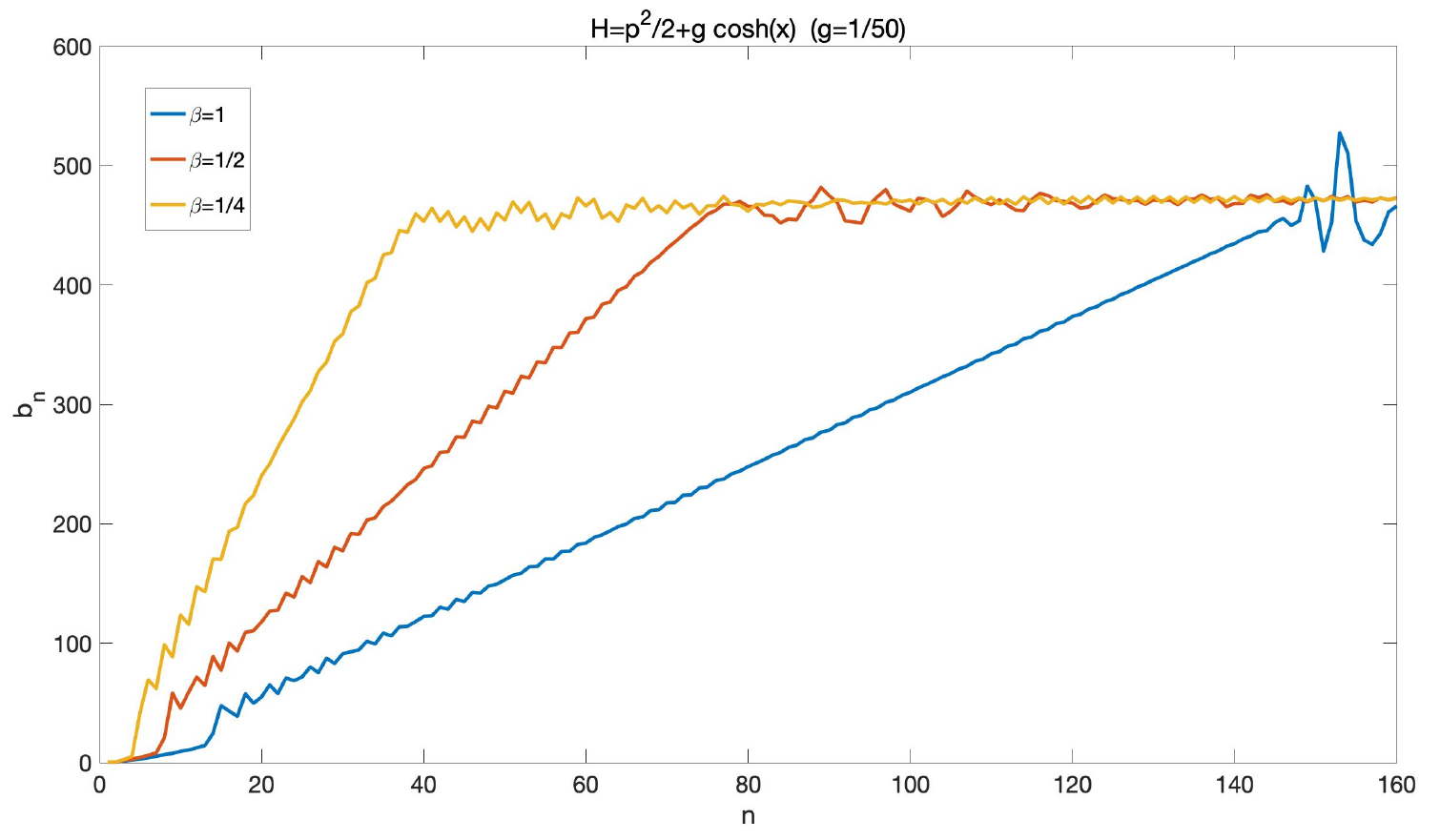} 	
	\end{subfigure}
	\begin{subfigure}{1\textwidth}
		\includegraphics[width=\textwidth]{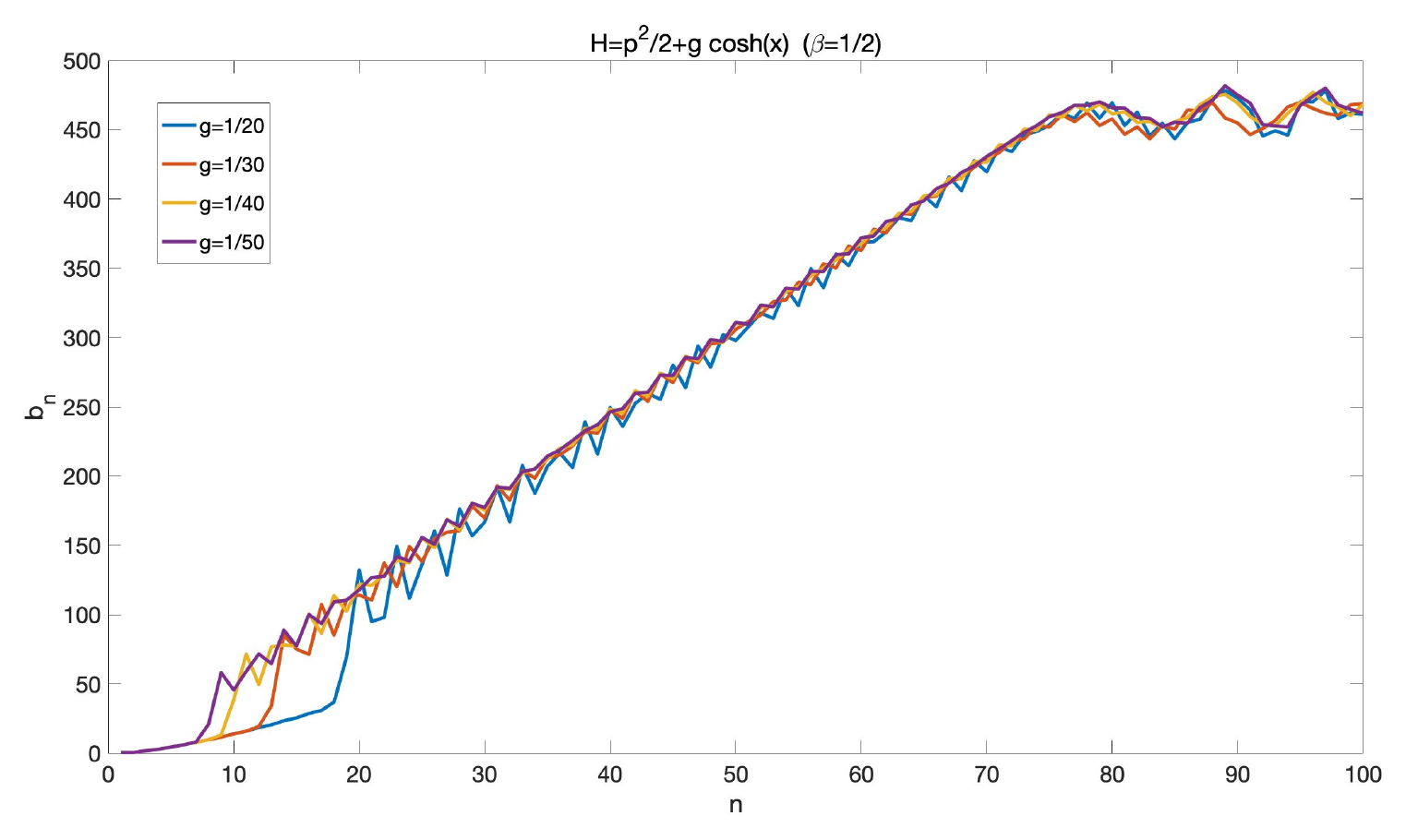}		
	\end{subfigure}
	\caption{The Lanczos coefficients for the non-relativistic Toda model}
	\label{mathieudiagram}
\end{figure*}

Similar to the previous sections, here we also choose initial operator $\mathcal{O}_0=\hat{x}$, with $\hbar=1$ and truncation dimension 250 and the coupling $g$ in the range of $\frac{1}{20}$ to $\frac{1}{50}$. In Figure \ref{mathieudiagram}, we plot the Lanczos coefficients , and find that for this integrable model, the figure is similar to the quartic anharmonic oscillator. At the beginning, the Lanczos coefficients grow linearly, then they reach the plateau.  As shown in the Table \ref{bnmathieu}, they also have the same property that the the slopes are proportional to $T$, and the ratios are less than but quite close to $\pi$, nearly saturating the bound (\ref{bound3.3}).

\begin{figure*}
	\centering
			\includegraphics[width=\textwidth]{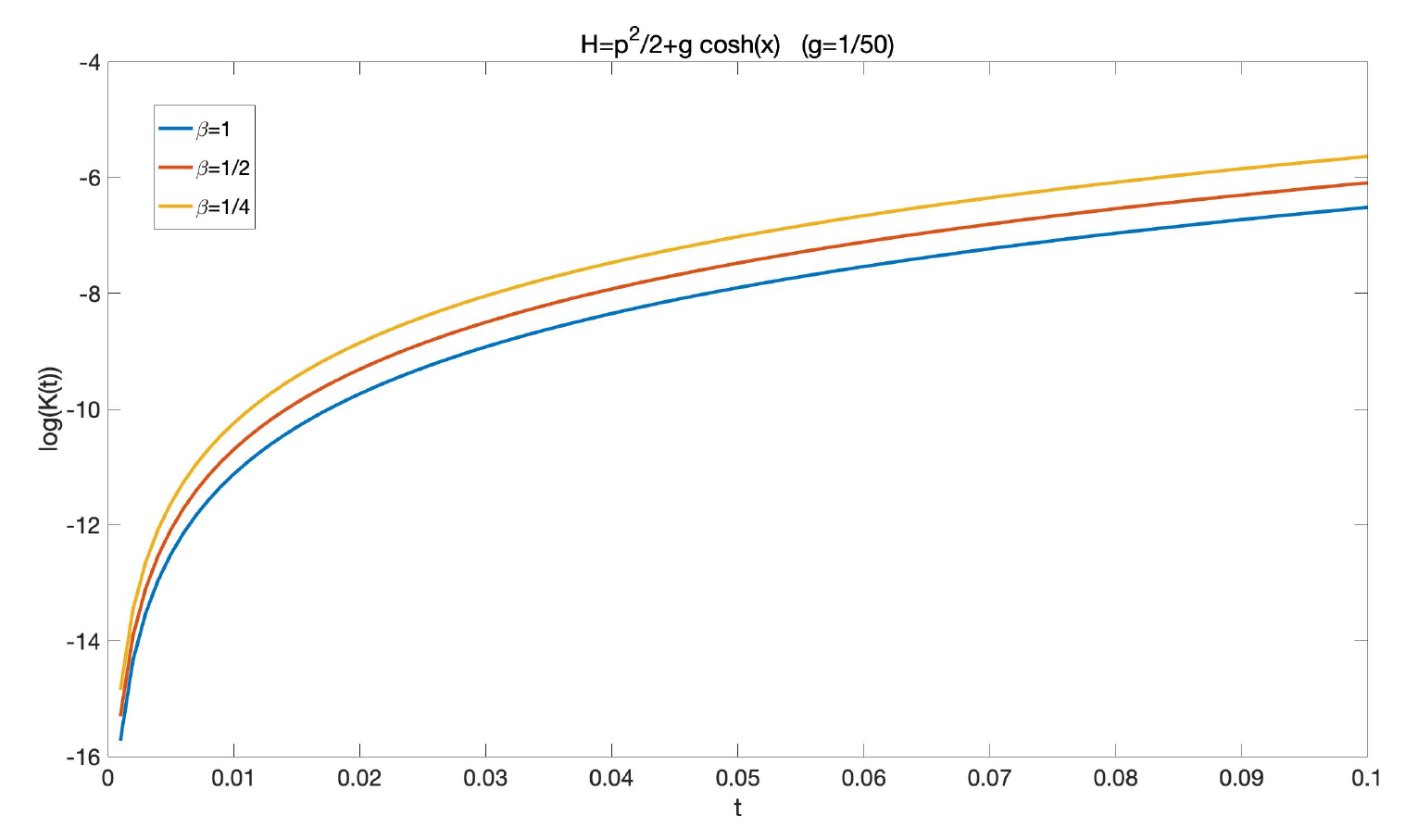} 	

	\caption{The Krylov complexity for the non-relativistic Toda model}
	\label{complexitymathieu}

\end{figure*}

In Figure \ref{complexitymathieu}, we plot the function $\log(K(t))$ for different $\beta$'s. Comparing with Figure \ref{complexityx4}  for the quartic anharmonic oscillator model in the previous section, both have a rapid growth in the small $t$ region then approach a slower and more linear growth. Apart from the small $t$ region, the graph here looks more linear, so the Krylov complexity for this model seems closer to exponential growth than that of the previous section.

We see that for this model, the properties the Lanczos coefficients and Krylov complexity are very similar to the quartic anharmonic oscillator in the previous section. The reason is that for the small coupling  $g$, one can expand the Hamiltonian 
	\be\ba
		H&= p^2+g \cosh(x)\\
		&=g+ p^2+ \frac{g}{2}x^2+\frac{g}{24}x^4+\dots,
	\ea\ee
where the terms of higher powers of $x$ in the potential  can be regard as small perturbations. After some proper rescaling, this can be indeed transformed into the quartic anharmonic oscillator.

\begin{table*}
\begin{center}

	\begin{tabular} {|c|c|c|c|} \hline  $\beta$  & $1$ & $\frac{1}{2}$ & $\frac{1}{4}$ \\
		 \hline  $\alpha$     & 3.0946  & 6.1188 & 11.9578 \\
		  \hline  $\alpha \beta$  &  3.0946  & 3.0594 & 2.9895\\
		 \hline
	\end{tabular}
	
	\caption{  The slope $\alpha$  for the non-relativistic Toda model} \label{bnmathieu}
\end{center}

\end{table*}

\section{Discussions}
\label{discussions}

In this paper, we study the Lanczos coefficients and Krylov complexity for a variety of different models using mostly numerical methods. For a linearly growing Lanczos coefficients, we also provide an alternative derivation on a bound on the slope by the temperature (\ref{bound3.3}) using the radius of convergence in the autocorrelation function.  

For the Calabi-Yau models in section \ref{CY}, we find that the Lanczos coefficients grow slower than linearly, and the local $\mathbb{P}^2$ model fits better with the power scaling law $b_n\sim n^\delta$ than the local $\mathbb{P}^1\times\mathbb{P}^1$ model. Their behaviors are stable with increasing truncation dimension in our numerical calculations. It seems that in this case, the scrambling effects contributed by saddles are minor comparing to the other models in the subsequent sections.  Meanwhile, we choose two different initial operators $\mathcal{O}_0=e^{\hat{x}},\hat{x}$, and find they have almost the same shape as Figure \ref{figurep1p1} and Figure \ref{figurep2} show, so the growth of Lanczos coefficients is not sensitive to the initial operators in this case. We numerically compute the best fit values for the scaling parameter $\delta$. 

There are certainly some possible future improvements of our analysis. As we mentioned,  the fits in Figure \ref{fit} are not overall quite good. It seems that the simple scaling function $n^\delta$ is not able to capture some fine details of the data.  The deviations from the scaling behavior may come from multiple competing physical mechanisms, including the contributions by saddles mentioned earlier. Further studies of these effects should provide a better template to fit the data. Furthermore,  although the effects of the truncation size on the Lanczos coefficients are smaller before reaching the plateau in Figures \ref{figurep1p1}, \ref{figurep2},  they are still quite visible in certain regimes, especially at low temperature. Currently it is not easy for us to further significantly increase the truncation size or accurately estimate the corrections.  In order to understand the fine features of the data, it is essential to also improve the numerical accuracy with better computational resource or more efficient algorithms. It would be also interesting to search for some analytic approaches to the problem.

On the other hand, for the non-relativistic models in sections \ref{quartic} and \ref{Toda}, we find that the Lanczos coefficients grow linearly at the beginning, and finally reach the plateau. And for our models, the slopes are proportional to $T$, with near saturation of the bound (\ref{bound3.3}), similar to e.g. the inverted harmonic oscillator studied in \cite{Baek:2022pkt}. The linear behavior is due to the saddle-dominated scrambling effects \cite{Xu:2019lhc,Bhattacharyya:2020art}, as we do not expect such simple one-body models to be truly chaotic. So in these cases the growth of  Lanczos coefficients does not provide a decisive good indicator of the integrable or chaotic nature of the system. It would be interesting to provide a physical explanation of the near saturation of the bound (\ref{bound3.3}). It would be interesting to have a more quantitative understanding  of the oscillation of Lanczos coefficients which is particularly evident in Section  \ref{quartic}. 

In this paper we consider the Krylov complexity defined from operator growth. It would be interesting to also consider the novel proposal of state complexity defined by minimizing the spread of a wave function \cite{Balasubramanian:2022tpr}. There are certainly many other proposed measures of complexity in the literature, e.g. as recently reviewed in context of the larger topic of quantum information in holographic duality in \cite{Chen:2021lnq}. One may also study them in the Calabi-Yau models.

As a first step, in this paper we only consider one-body quantum dynamics. It would be interesting to couple the quantum systems together and  study the many-body quantum dynamics. It would be certainly interesting to explore the connections to holographic duality.

\vspace{0.2in} {\leftline {\bf Acknowledgments}}
\nopagebreak

We thank Jun-hao Li, Gao-fu Ren, Wen-qi Yu, Pei-xuan Zeng for helpful discussions. This work was supported in parts by the national Natural Science Foundation of China (Grants  No.11947301 and No.12047502).

\addcontentsline{toc}{section}{References}


\providecommand{\href}[2]{#2}\begingroup\raggedright\endgroup

\end{document}